\documentclass[aps,preprint,nofootinbib]{revtex4}
\usepackage{natbib}
\usepackage[]{fontenc}
\usepackage[latin1]{inputenc}
\usepackage{amsmath}
\usepackage{graphicx}
\usepackage[colorlinks=true,linkcolor=blue,citecolor=blue,filecolor=blue,urlcolor=blue]{hyperref}

\makeatletter

\providecommand{\tabularnewline}{\\}

\usepackage[english]{babel}

\newcommand{\comment}[1]{ }

\newcommand{\half}{\frac{1}{2}}

\newcommand{\phd}{\delta}
\newcommand{\phm}{\phi}

\newcommand{\fmt}{f_{\mu\tau}}
\newcommand{\gee}{g_{ee}}
\newcommand{\gem}{g_{e\mu}}
\newcommand{\get}{g_{e\tau}}

\makeatother
\begin{document}

\preprint{FTUV-14-0214, IFIC/14-13\\}

\title{The Zee-Babu Model revisited in the light of new data}

\author{Juan Herrero-Garcia$^a$, Miguel Nebot$^b$, Nuria Rius$^a$ and Arcadi Santamaria$^a$}

\affiliation{$^a$ Departament de Física Teòrica, Universitat de València and\\
IFIC, Universitat de València-CSIC\\
Dr. Moliner 50, E-46100 Burjassot (València), Spain}
\affiliation{$^b$ Centro de Física Teórica de Partículas,\\ Instituto Superior Técnico -- Universidade de Lisboa\\ Av. Rovisco Pais 1, 1049-001 Lisboa, Portugal}

\begin{abstract}

\noindent We update previous analyses of the Zee-Babu model in the light of new data, e.g., the mixing angle $\theta_{13}$, the rare decay $\mu\to e \gamma$ and the LHC results. We also analyse the possibility of accommodating the deviations in $\Gamma(H\to \gamma\gamma)$ hinted by the LHC experiments, and the stability of the scalar potential.  
We find that neutrino oscillation data and low energy constraints are still compatible with masses of the extra charged scalars accessible to LHC.
Moreover, if any of them is discovered, the model can be 
falsified by combining the information on the singly and doubly charged scalar decay modes with neutrino data. Conversely, 
if the neutrino spectrum is found to be inverted and the CP phase $\delta$  is quite different from $\pi$, the masses of the
charged scalars will be well outside the LHC reach.

\end{abstract}
\maketitle

\section{Introduction}

The observed pattern of neutrino masses and mixing remains  one of the major puzzles in particle physics. Moreover,  massive neutrinos provide irrefutable evidence for physics beyond the Standard Model (SM) and many theoretical possibilities have been proposed to account for  the lightness of neutrinos (see \cite{Mohapatra:2005wg,GonzalezGarcia:2007ib,Beringer:1900zz,deGouvea:2013onf} for some reviews). 
With the running of the LHC, it is timely to explore neutrino mass models in which the scale of new physics is close to the TeV. In particular, radiative mechanisms are especially appealing, since 
small neutrino masses are generated naturally due to loop factors. On the other hand, new physics effects can be sizable also in low energy experiments, for instance lepton
flavour violating rare decays of charged leptons, 
$\ell_\alpha \rightarrow \ell_\beta \gamma$, providing complementary probes for such  
models.

In this paper we consider 
the Zee-Babu model (ZB) of neutrino masses\footnote{
The model was first proposed in \cite{Zee:1985id} and studied carefully in \cite{Babu:1988ki}. Similar models with a doubly charged scalar and masses generated at two loops were discussed in  \cite{Cheng:1980qt} (two-loop neutrino mass models containing doubly-charged singlets have also been recently discussed in connection with neutrinoless double beta decay \cite{delAguila:2011gr,delAguila:2013zba}).},  which just adds two (singly and doubly) charged scalar singlets to the SM. 
Neutrino masses are generated at two loops and  are proportional to the
Yukawa couplings of the new scalars and inversely proportional to
the square of their masses. This is phenomenologically quite interesting because
the new scalars cannot be very heavy or have very small Yukawa couplings, otherwise neutrino masses would be too small.
As a consequence, such scalars may be accessible at the LHC, and in principle they could explain the slight excess over the SM prediction found by ATLAS in  the
diphoton Higgs decay channel $H\rightarrow \gamma \gamma$
(currently CMS does not see any excess, see section \ref{sec:gamma} for the latest data).
They also mediate a variety of  lepton flavour violating (LFV) processes,  
leading to rates measurable in current experiments.

The phenomenology of the ZB model has been widely analyzed:
neutrino oscillation data was used to constrain the parameter space of the model,  LFV 
charged lepton decay rates calculated and collider signals discussed
\cite{Babu:2002uu,AristizabalSierra:2006gb,Nebot:2007bc}. 
Non-standard neutrino interactions in the ZB model have also been thoroughly studied,  
in correlation with possible LHC signals and LFV processes \cite{Ohlsson:2009vk}.
 In  \cite{Nebot:2007bc}, some of us performed an exhaustive 
numerical study of the full parameter space of the model using Monte Carlo Markov Chain 
(MCMC) techniques, which allow to efficiently explore high-dimensional spaces. 
However,  in the last few years  there have been 
several experimental results which  
motivate an up-to-date analysis including all relevant data currently available.
Therefore, in this work 
we update previous analysis in the light of the recent measurement of the neutrino mixing angle 
$\theta_{13}$ 
\cite{Abe:2011fz,An:2012eh,Ahn:2012nd}, the new MEG limits 
on $\mu\rightarrow e \gamma$ \cite{Adam:2013mnn}, the lower bounds on doubly-charged scalars coming from LHC data \cite{ATLAS:2012hi,Chatrchyan:2012ya},
and, of course, the discovery of a 125 GeV Higgs boson by ATLAS and CMS \cite{Aad:2012tfa,Chatrchyan:2012ufa}.
Moreover, we 
also study the possibility of accommodating deviations from the SM prediction for the Higgs diphoton decay channel, and the effects of the new couplings of the model in the stability of the scalar potential. A possible  enhancement of the  Higgs diphoton decay rate in the ZB model together with the vacuum
stability of the scalar potential has been studied in \cite{Chao:2012xt}, however a consistent 
updated analysis including all constraints is lacking.

The outline of the paper is the following. In section \ref{sec:ZB-model} we briefly review the main features of the ZB model, discussing perturbativity and naturality 
estimates for the allowed ranges of the free parameters of the model. 
We summarize present constraints from  recent neutrino oscillation data, 
low energy lepton-flavour violating processes, universality and stability of the scalar potential. 
We also review the collider phenomenology of the ZB model, discussing current limits from LHC,
and briefly comment on the prospects for non-standard neutrino interactions.  
In section \ref{sec:gamma} we 
analyze in detail the contributions of the ZB charged scalars to both, $\Gamma(H \rightarrow \gamma \gamma)$ and $\Gamma(H \rightarrow Z \gamma)$.
After some analytic estimates in section \ref{sec:ae},  we present the results of our MCMC numerical analysis in section \ref{sec:Numerical-analysis} and we conclude in section \ref{sec:Conclusions}. Renormalization group equations for the ZB model and relevant loop functions 
are collected in the appendices.

\section{The Zee-Babu model \label{sec:ZB-model}}

We  follow the notation of \cite{Nebot:2007bc}. 
As mentioned above, the Zee-Babu model only contains, in addition to the SM, two charged singlet scalar fields \begin{equation}
h^{\pm},\qquad k^{\pm\pm}\,,\end{equation}
with weak hypercharges $\pm 1$ and $\pm 2$ respectively (we use 
 the convention $Q=T_{3}+Y$).

The scalar potential is given by
 \begin{eqnarray}
V & = & m_{H}^{\prime2} H^{\dagger}H+ m_{h}^{\prime2}|h|^{2}+m_{k}^{\prime2}|k|^{2} + 
\lambda_{H}(H^{\dagger}H)^{2}+\lambda_{h}|h|^{4}+\lambda_{k}|k|^{4} 
\nonumber \\
&+&\lambda_{hk}|h|^{2}|k|^{2} +  \lambda_{hH}|h|^{2}H^{\dagger}H+\lambda_{kH}|k|^{2}H^{\dagger}H+\left(\mu h^{2}k^{++}+\mathrm{h.c.}\right)\,, 
\label{eq:V}
\end{eqnarray}
being $H$ the $SU(2)$ doublet Higgs boson, while 
the leptons have Yukawa couplings to both  $H$ and 
the new charged scalars:
\begin{equation}
\mathcal{L}_{Y}=
\overline{L_L}\, Y e H +  \overline{\tilde{L}_L}f\ell h^{+}+\overline{e^{c}}g\, e\, k^{++}+
\mathrm{h.c.}  \,, \label{eq:babuyuk}
\end{equation}
where $L_L$ and $e$ are the SM $SU(2)$ lepton doublets and singlets,  
respectively, 
and $\tilde{L}_L \equiv i \tau_2 L_L^c = i \tau_2 C \overline{L_L}^T$, with $\tau_2$ Pauli's second 
matrix. 
Due to Fermi statistics, $f_{ab}$ is an antisymmetric matrix in flavour space while $g_{ab}$ is symmetric.

Notice that we can assign lepton number $-2$
to both scalars, $h^+$ and $k^{++}$, in such a way that total lepton
number  $L$  (or $B-L$) is conserved in 
 the complete Lagrangian, except for the trilinear coupling $\mu$ of the scalar potential; 
thus, lepton number is explicitly broken by the 
$\mu$-coupling. 
It is important to remark that lepton number violation requires the simultaneous
presence of the four couplings $Y$, $f$, $g$ and $\mu$, because
if any of them vanishes one can always assign quantum numbers in such
a way that there is a global $U(1)$ symmetry. This means that neutrino
masses will require the simultaneous presence of the four couplings. 

Regarding the  physical free parameters in the ZB model, our convention is the following: 
without loss of generality, we choose the 3 $\times$ 3 charged lepton Yukawa matrix $Y$ 
to be diagonal with real and positive elements. We also use fermion field rephasings to remove three phases from the elements of the matrix $g$  and  charged scalar rephasings 
to set $\mu$ real and positive, and to remove one  phase from $f$. 
In summary we have 
 $12$ moduli ($3$ from $Y$, $3$ from $f$ and $6$ from $g_{ab}$), $5$ phases ($3$ from 
 $g$ and $2$ from $f$) and the real and positive parameter $\mu$, plus the rest 
 of real parameters in the scalar potential. 
As discussed in \cite{Nebot:2007bc},  this choice is compatible with the standard parametrization of neutrino masses and mixings. 

After electroweak symmetry breaking, 
the masses of charged leptons are $m_{a}=Y_{aa}v$, with 
$v\equiv\left\langle H^{0}\right\rangle =174\,\mathrm{GeV}$, the VEV of the standard Higgs doublet, while the physical charged scalar masses are given by 
\begin{equation}
m_{h}^{2}=m_{h}^{\prime2}+\lambda_{hH}v^{2}\:,\qquad m_{k}^{2}=m_{k}^{\prime2}+\lambda_{kH}v^{2}\,.
\end{equation}

In principle, the scale of the new mass parameters of the ZB model 
($m_{h}, m_{k}$ and $\mu$) is arbitrary. However  
 from the experimental point of view it is interesting to consider 
new scalars light enough to be produced in the second run of the LHC. Also 
theoretical arguments suggest that the scalar masses should be relatively 
light (few TeV), to avoid unnaturally large one-loop corrections to the Higgs mass
which would introduce a hierarchy problem.
Therefore, in this paper we will focus on the masses of the new scalars, 
$m_{h}, m_{k}$, below 2 TeV.

The Yukawa couplings of the new scalars of the model enter in 
the neutrino mass formula and in several LFV processes, and are strongly bounded for the scalar masses we are considering except in a few corners of the parameter space where we require that the theory remains perturbative. Since one-loop corrections to Yukawa couplings are order
\begin{equation}
\delta f\sim\frac{f^{3}}{(4\pi)^{2}}\,,\qquad\delta g\sim\frac{g^{3}}{(4\pi)^{2}}\,.
\end{equation}
one expects from perturbativity $f,g \ll 4\pi$, although, as we will see, for the scalar masses considered here, phenomenological constraints are always stronger.

The couplings of the charged scalars in the scalar potential, apart from the stability constraints described in section \ref{sub:Stability}, are essentially free. However, for the theory to make sense as a perturbative theory we also impose the limit\footnote{Notice that there could be order one differences in the perturbativity constraints on the different couplings $\lambda_i$ from perturbative unitarity of the matrix elements \cite{Dicus:1992vj,Lee:1977eg}. We can neglect them for the purpose of this work, keeping in mind that they could be relevant when perturbativity is ``pushed" to the limit (as needed to explain $H\rightarrow \gamma \gamma$ enhancement, see sec.~\ref{sec:gamma}).} $\lambda_{h,k,kH,hH,hk} < 4 \pi$.
   
The trilinear coupling among charged scalars $\mu$, on the other hand, is different, for it has dimensions of mass and it is insensitive to high energy perturbative unitarity constraints. However, it induces radiative corrections to the masses of the charged scalars of order 
\begin{equation}
\delta m_{k}^{2},\delta m_{h}^{2}\sim\frac{\mu^{2}}{(4\pi)^{2}}\,.
\end{equation}
Requiring that the corrections in absolute value are much smaller than the masses 
we can derive a naive upper bound for this parameter, $\mu\ll 4\pi \,  {\rm min}(m_{h},m_k)$, but it is difficult to fix an exact value of $\mu$ for which the contributions to the scalar masses are unacceptably large, leading to a highly fine-tuned scenario.  

A large value of $\mu$, as compared with scalar masses, is also disfavoured because it could lead to a deeper minimum of the scalar potential for non-vanishing values of the charged fields, therefore breaking charge conservation. This phenomenon has also been studied in the context of supersymmetric theories (see for instance \cite{Frere:1983ag,AlvarezGaume:1983gj,Casas:1996de}).  As an example,  by looking at the particular direction $|H|=|h|=|k|=r$, and 
requiring that the charge breaking minimum  is not  a global minimum, $V(r \neq 0) > 0$, 
one  obtains
\begin{equation}
 \mu^2 < \left(\lambda_H+\lambda_h+\lambda_k+\lambda_{hH} + \lambda_{kH}+\lambda_{hk}\right)\left(m_H^{\prime 2}+m_h^{\prime 2}+m_k^{\prime 2}\right) . 
\end{equation}
Assuming no cancellations between the $\lambda$'s or mass terms, neglecting $\lambda_H$ and $m_H^{\prime 2}$, and using the perturbative limit for the rest of the couplings $\lambda_i\lesssim 4\pi$ one finds a very conservative bound on $\mu$
\begin{equation}
\mu\lesssim \sqrt{20\pi} \max(m_k,m_h) \sim 8 \max(m_k,m_h)
\end{equation}
Tighter limits can be obtained by looking at all directions in the potential and/or allowing for cancellations.

Given that the neutrino masses depend linearly on the parameter $\mu$, as we will see in the next section, the ability of the model to accommodate all present data is  quite sensitive to the upper limit allowed for $\mu$. 
Thus we choose to implement such limit in terms of a parameter $\kappa$, 
\begin{equation} \label{mu}
\mu< \kappa \, {\rm min} (m_{h},m_k)  \,,
\end{equation}
and discuss our results for different values of $\kappa=1,5,4\pi$. Notice that we are using the naturality upper bound (expressed in terms of ${\rm min} (m_{h},m_k)$), which 
 in general is much
more restrictive than the upper bound obtained by requiring that the minimum of the potential does not break charge conservation (expressed in terms of  ${\rm max} (m_{h},m_k)$).

\subsection{Neutrino masses. }

\label{secmas} 
\begin{figure}
\begin{centering}\includegraphics[width=0.6\columnwidth]{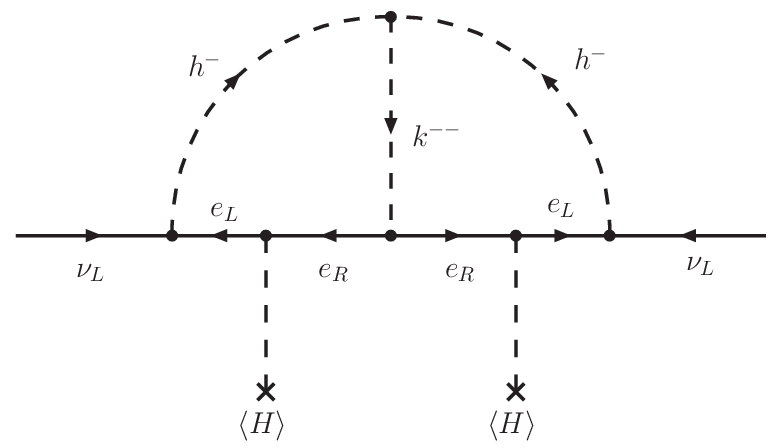}\par\end{centering}

\caption{Diagram contributing to the neutrino Majorana mass at two loops.}

\label{fig:babumass}
\end{figure}

The lowest order contribution to neutrino masses involving
the four relevant couplings appears at two loops \cite{Zee:1985id,Babu:1988ki}
and its Feynman diagram is depicted in fig.~\ref{fig:babumass}.

The calculation of this diagram gives the following mass matrix for
the neutrinos (defined as an effective term in the Lagrangian $\mathcal{L}_{\nu}\equiv-\half\overline{\nu_{L}^{c}}\mathcal{M}_{\nu}\nu_{L}+\mathrm{h.c.}$) 
\begin{equation}
(\mathcal{M}_{\nu})_{ij}=16\mu f_{ia}m_{a}g_{ab}^{*}I_{ab}m_{b}f_{jb}\,,
\label{eq:babumas}\end{equation}
 where
$I_{ab}$ is the two-loop integral, which 
can be calculated analytically \cite{McDonald:2003zj}.
However, since $m_{c},m_{d}$ are the masses of the charged leptons,
necessarily much lighter than the charged scalars, we can neglect
them and obtain a much simpler form\begin{eqnarray}
I_{cd} & \simeq & I=\frac{1}{(16\pi^{2})^{2}}\frac{1}{M^{2}}\frac{\pi^{2}}{3}\tilde{I}(r)\quad,\quad M\equiv\max(m_{h},m_{k})\,,\label{eq:int_mas}\end{eqnarray}
where $\tilde{I}(r)$ is a function of the ratio of the masses of
the scalars $r\equiv m_{k}^{2}/m_{h}^{2}$, 
\begin{equation}
\tilde{I}(r)=\begin{cases}
1+\frac{3}{\pi^{2}}(\log^{2}r-1) & \text{for }r\gg1\\
1 & \text{for }r\rightarrow0\end{cases}\,,\end{equation}
which is close to one for a wide range of scalar masses. Within this
approximation the neutrino mass matrix can be directly written in
terms of the Yukawa coupling matrices, $f$, $g$, and $Y$ \begin{equation}
\mathcal{M}_{\nu}=\frac{v^{2}\mu}{48\pi^{2}M^{2}}\tilde{I}\, f\, Y\, g^{\dagger}Y^{T}f^{T}\,.\label{eq:MnuYukawas}\end{equation}
A very important point is that since $f$ is a $3\times 3$ antisymmetric
matrix, $\det f=0$ (for $3$ generations), and therefore $\det\mathcal{M}_{\nu}=0$. Thus,
at least one of the neutrinos is exactly massless at this order.

The neutrino Majorana mass matrix $\mathcal{M}_{\nu}$ can be written as 
\begin{equation}
\mathcal{M}_{\nu}= U D_\nu U^T \,, 
\label{eq:numass}
\end{equation}
where $D_\nu$ is a diagonal  matrix with real positive eigenvalues, 
 and $U$ is the 
 Pontecorvo-Maki-Nakagawa-Sakata (PMNS)  leptonic mixing matrix. 
 We are  left with only  two possibilities for the neutrino masses, $m_i$:

\begin{itemize}
\item Normal hierarchy (NH):  the solar squared mass difference is
$\Delta_S = m_2^2$, the atmospheric mass splitting 
 $\Delta_A = m_3^2$ and $m_1=0$, with $m_3 \gg m_2$ . 
\item Inverted hierarchy (IH):  $\Delta_S = m_2^2-m_1^2$,  
 $\Delta_A = m_1^2$ and $m_3=0$, with $m_1 \approx  m_2$.
\end{itemize}

The standard parametrization for the PMNS matrix is
 \begin{eqnarray}
 U=\left(\begin{array}{ccc}
c_{13}c_{12} & c_{13}s_{12} & s_{13}e^{-i\delta}\\
-c_{23}s_{12}-s_{23}s_{13}c_{12}e^{i\delta} & c_{23}c_{12}-s_{23}s_{13}s_{12}e^{i\delta} & s_{23}c_{13}\\
s_{23}s_{12}-c_{23}s_{13}c_{12}e^{i\delta} & -s_{23}c_{12}-c_{23}s_{13}s_{12}e^{i\delta} & c_{23}c_{13}\end{array}\right)\left(\begin{array}{ccc}
1 \\
 & e^{i\phi/2}\\
 &  & 1\end{array}\right)\ ,\label{UPMNS}
 \end{eqnarray}
where $c_{ij}\equiv\cos\theta_{ij}$, $s_{ij}\equiv\sin\theta_{ij}$ and since one of the neutrinos is massless, there is only one physical Majorana phase, $\phi$, in addition to the Dirac phase 
$\delta$.

\subsection{Low energy constraints. \label{sub:Low-energy-constraints} }

In order to provide neutrino masses compatible with experiment, the
Yukawa couplings of the charged scalars cannot be too small and their
masses cannot be too large. This immediately gives rise to a series
of flavour lepton number violating processes, as for instance $\mu^{-}\rightarrow e^{-}\gamma$
or $\mu^{-}\rightarrow e^{+}e^{-}e^{-}$, with rates which can be,
in some cases, at the verge of the present experimental limits. Therefore,
 we can use these processes to obtain information about the
parameters of the model and hopefully to confirm or to exclude the model
in a near future by exploiting the  synergies with direct searches for the new scalars at LHC.

 In this section we  follow the notation of \cite{Nebot:2007bc}, where all the relevant formulae 
 can be found, and update the new bounds. 
 We collect the relevant tree-level lepton flavour violating constraints, 
 from $\ell^-_a \rightarrow \ell^+_b  \ell^-_c  \ell^-_d$ decays and 
 $\mu^+ e^- \leftrightarrow \mu^- e^+$ transitions, in table \ref{tab:meee}. 

\begin{table}
\begin{tabular}{ccl}
\hline 
Process&
Experiment (90\% CL)&
Bound (90\% CL)\tabularnewline
\hline
$\mu^{-}\rightarrow e^{+}e^{-}e^{-}$&
BR$<1.0\times10^{-12}$&
$|g_{e\mu}g_{ee}^{*}|<2.3\times10^{-5}\,\left(\frac{m_{k}}{\mathrm{TeV}} \right)^{2}$\tabularnewline
$\tau^{-}\rightarrow e^{+}e^{-}e^{-}$&
BR$<2.7\times10^{-8}$&
$|g_{e\tau}g_{ee}^{*}|<0.009\,\left(\frac{m_{k}}{\mathrm{TeV}} \right)^{2}$\tabularnewline
$\tau^{-}\rightarrow e^{+}e^{-}\mu^{-}$&
BR$<1.8\times10^{-8}$&
$|g_{e\tau}g_{e\mu}^{*}|<0.005\,\left(\frac{m_{k}}{\mathrm{TeV}} \right)^{2}$\tabularnewline
$\tau^{-}\rightarrow e^{+}\mu^{-}\mu^{-}$&
BR$<1.7\times10^{-8}$&
$|g_{e\tau}g_{\mu\mu}^{*}|<0.007\,\left(\frac{m_{k}}{\mathrm{TeV}} \right)^{2}$\tabularnewline
$\tau^{-}\rightarrow\mu^{+}e^{-}e^{-}$&
BR$<1.5\times10^{-8}$&
$|g_{\mu\tau}g_{ee}^{*}|<0.007\,\left(\frac{m_{k}}{\mathrm{TeV}} \right)^{2}$\tabularnewline
$\tau^{-}\rightarrow\mu^{+}e^{-}\mu^{-}$&
BR$<2.7\times10^{-8}$&
$|g_{\mu\tau}g_{e\mu}^{*}|<0.007\,\left(\frac{m_{k}}{\mathrm{TeV}} \right)^{2}$\tabularnewline
$\tau^{-}\rightarrow\mu^{+}\mu^{-}\mu^{-}$&
BR$<2.1\times10^{-8}$&
$|g_{\mu\tau}g_{\mu\mu}^{*}|<0.008\,\left(\frac{m_{k}}{\mathrm{TeV}} \right)^{2}$\tabularnewline
$\mu^{+}e^{-}\rightarrow\mu^{-}e^{+}$&
$G_{M\bar{M}}<0.003G_{F}$&
$|g_{ee}g_{\mu\mu}^{*}|<0.2\,\left(\frac{m_{k}}{\mathrm{TeV}} \right)^{2}$\tabularnewline
\hline
\end{tabular}
\caption{Constraints from tree-level lepton flavour violating decays~\cite{Beringer:1900zz}.\label{tab:meee}}
\end{table}
 
Universality constraints are summarized in table \ref{tab:universality} where we have combined the measurements presented in \cite{Pich:2013lsa}  for the different couplings. There seems to be a $2\sigma$ discrepancy in $G_{\tau}^{exp}/G_{e}^{exp}$, which we interpret as a bound. If confirmed and interpreted within the ZB model, one obtains that $|f_{\mu\tau}|^{2}-|f_{e\mu}|^{2}=0.05 \,(m_{h}/\mathrm{TeV})^{2}$. 
As we will see in section~\ref{sec:ae}, for NH spectrum $f_{e\mu} \sim f_{\mu\tau}/2$, therefore 
one needs $m_h\sim 4\,f_{\mu\tau}$TeV, which is easily achieved. For IH spectrum, however,  $f_{\mu\tau}\sim 0.2 f_{e\mu}$ ($f_{\mu\tau}\sim (0.15-0.3)\,f_{e\mu}$ if we vary the angles in their $3\sigma$ range), and therefore, if this measurement is confirmed, the IH scheme in the ZB model would be disfavoured. 

\begin{table}
\begin{tabular}{ccc}
\hline 
SM Test&
Experiment&
Bound (90\%CL)\tabularnewline
\hline
lept./hadr. univ.&
$\sum_{q=d,s,b}|V_{uq}^{exp}|^{2}=0.9999\pm0.0006$&
$|f_{e\mu}|^{2}<0.007\,\left(\frac{m_{h}}{\mathrm{TeV}} \right)^{2}$\tabularnewline
$\mu/e$ universality&
$\frac{G_{\mu}^{exp}}{G_{e}^{exp}}=1.0010\pm0.0009$&
$||f_{\mu\tau}|^{2}-|f_{e\tau}|^{2}| <0.024\,\left(\frac{m_{h}}{\mathrm{TeV}} \right)^{2}$\tabularnewline
$\tau/\mu$ universality&
$\frac{G_{\tau}^{exp}}{G_{\mu}^{exp}}=0.9998\pm0.0013$&
$||f_{e\tau}|^{2}-|f_{e\mu}|^{2}|<0.035\,\left(\frac{m_{h}}{\mathrm{TeV}} \right)^{2}$\tabularnewline

$\tau/e$ universality &
$\frac{G_{\tau}^{exp}}{G_{e}^{exp}}=1.0034\pm0.0015$&
$||f_{\mu\tau}|^{2}-|f_{e\mu}|^{2}|<0.04\,\left(\frac{m_{h}}{\mathrm{TeV}} \right)^{2}$\tabularnewline
\hline
\end{tabular}
\caption{Constraints from universality of charged currents obtained combining the experimental results compiled in table 2 of
 \cite{Pich:2013lsa}.\label{tab:universality}}
\end{table}

Finally, one-loop level 
lepton flavour violating constraints coming from $\ell^-_a \rightarrow  \ell^-_b \gamma$  decays\footnote{As was shown in \cite{Raidal:1997hq}, 
doubly charged scalars can give logarithmic enhanced contributions to muon-electron conversion in nuclei. 
Moreover, planned experiments will improve current limits by four orders of magnitude
\cite{Witte:2012zza,Kuno:2013mha,Onorato:2013uka}; however, at present, limits are still not competitive with $\mu\rightarrow e \gamma$.} 
and anomalous magnetic moments of electron and muon are collected
in table \ref{tab:meg}, including the recent limit on BR($\mu\rightarrow e \gamma$)
from the MEG Collaboration \cite{Adam:2013mnn}. 
 
\begin{table}
\begin{tabular}{cl}
\hline 
Experiment&
~~~~~Bound (90\%CL)\tabularnewline
\hline
$\delta a_{e}=(12\pm10)\times10^{-12}$&
$r\left(|f_{e\mu}|^{2}+|f_{e\tau}|^{2}\right)+4\left(|g_{ee}|^{2}+|g_{e\mu}|^{2}+|g_{e\tau}|^{2}\right)<5.5\times10^{3}\,(m_{k}/\mathrm{TeV})^{2}$\tabularnewline
$\delta a_{\mu}=(21\pm10)\times10^{-10}$&
$r\left(|f_{e\mu}|^{2}+|f_{\mu\tau}|^{2}\right)+4\left(|g_{e\mu}|^{2}+|g_{\mu\mu}|^{2}+|g_{\mu\tau}|^{2}\right)<7.9\,(m_{k}/\mathrm{TeV})^{2}$\tabularnewline
$BR(\mu\rightarrow e\gamma)<5.7\times10^{-13}$&
$r^{2}|f_{e\tau}^{*}f_{\mu\tau}|^{2}+16|g_{ee}^{*}g_{e\mu}+g_{e\mu}^{*}g_{\mu\mu}+g_{e\tau}^{*}g_{\mu\tau}|^{2}<1.6\times10^{-6}\,(m_{k}/\mathrm{TeV})^{4}$\tabularnewline
$BR(\tau\rightarrow e\gamma)<3.3\times10^{-8}$&
$r^{2}|f_{e\mu}^{*}f_{\mu\tau}|^{2}+16|g_{ee}^{*}g_{e\tau}+g_{e\mu}^{*}g_{\mu\tau}+g_{e\tau}^{*}g_{\tau\tau}|^{2}<0.52\,(m_{k}/\mathrm{TeV})^{4}$\tabularnewline
$BR(\tau\rightarrow\mu\gamma)<4.4\times10^{-8}$&
$r^{2}|f_{e\mu}^{*}f_{e\tau}|^{2}+16|g_{e\mu}^{*}g_{e\tau}+g_{\mu\mu}^{*}g_{\mu\tau}+g_{\mu\tau}^{*}g_{\tau\tau}|^{2}<0.7\,(m_{k}/\mathrm{TeV})^{4}$\tabularnewline
\hline
\end{tabular}

\caption{Constraints from loop-level lepton flavour violating interactions and anomalous magnetic moments \cite{Beringer:1900zz,Adam:2013mnn}.\label{tab:meg}}
\end{table}

Given that lepton number is not conserved,  another interesting low energy
process that could arise in the ZB model is neutrinoless double beta
decay ($0\nu2\beta$). 
However, since the singly and doubly charged scalars do not couple to hadrons and are singlet under the weak SU(2) (therefore, do not couple to $W$ 
gauge bosons), the $0\nu2\beta$ rate is dominated by the Majorana neutrino exchange \cite{delAguila:2012nu} and it is
proportional to the $|(\mathcal{M}_{\nu})_{ee}|^{2}$ matrix element.
In the NH case, 
\begin{equation}
(\mathcal{M}_{\nu}^{NH})_{ee}=\sqrt{\Delta_{S}}c_{13}^{2}s_{12}^{2}e^{i\phi}+\sqrt{\Delta_{A}}s_{13}^{2}
\,.
\label{eq:2betaNH}
\end{equation}

Using neutrino oscillation data, one obtains  
$0.001  \lesssim \,\mathrm{eV} |(\mathcal{M}_{\nu}^{NH})_{ee}| \lesssim 0.004\,\mathrm{eV}$
and therefore it is outside the reach of present and near future $0\nu2\beta$ decay experiments.

In the IH case,  
\begin{equation}
(\mathcal{M}_{\nu}^{IH})_{ee}=\sqrt{\Delta_{A}+\Delta_{S}}c_{13}^{2}s_{12}^{2}e^{i\phi}+\sqrt{\Delta_{A}}c_{13}^{2}c_{12}^{2}\,.
\label{eq:2betaIH}\end{equation}
Then,
$0.01\,\mathrm{eV}  \lesssim |(\mathcal{M}_{\nu}^{NH})_{ee}| \lesssim 0.05\,\mathrm{eV}$
and, therefore, it is observable in planned $0\nu2\beta$
decay experiments.

\subsection{Non-standard interactions.\label{sub:NSI}}

The heavy scalars of the ZB model induce non-standard lepton interactions at tree level, 
which have been thoroughly analyzed in \cite{Ohlsson:2009vk}. In particular, by integrating out the singly charged scalar $h^+$,  the
following dimension-6 operators are generated:
\begin{equation}
\mathcal{L}_{d=6}^{NSI} = 2 \sqrt{2} G_F \epsilon_{\alpha \beta}^{\rho \sigma}
(\overline{\nu_{\alpha}} \gamma^\mu P_L \nu_\beta)
(\overline{\ell_{\rho}} \gamma_\mu P_L \ell_{\sigma}), 
\label{eq:nsi}
\end{equation}
where $\ell$ refer to the charged leptons and the standard NSI parameters $\epsilon_{\alpha \beta}^{\rho \sigma}$ are given by 
\begin{equation}
\epsilon_{\alpha \beta}^{\rho \sigma} = \frac{f_{\sigma\beta} f^*_{\rho\alpha}}{\sqrt{2} G_F m_h^2}.
\label{eq:eps}
\end{equation}
Regarding  neutrino propagation in matter, the relevant NSI parameters are 
$\epsilon^m_{\alpha \beta} = \epsilon^{ee}_{\alpha \beta}$. 
Since the couplings $f_{\sigma\beta}$ are antisymmetric, in the ZB model only 
$\epsilon^m_{\mu\tau}, \epsilon^m_{\mu\mu}$ and $\epsilon^m_{\tau \tau}$ are non zero. 

\noindent NSI can also affect the neutrino production in a neutrino factory,  via the processes
 $\mu\rightarrow e \overline{\nu_\beta} \nu_\alpha$. Source effects in the
 $\nu_\mu\rightarrow \nu_\tau$ and $\nu_e \rightarrow \nu_\tau$ channels are 
 produced by the NSI parameters 
 \begin{eqnarray}
 \epsilon^s_{\mu\tau} & = \epsilon^{e\mu}_{ \tau e} & = 
 \frac{f_{\mu e} f^*_{e \tau}}{\sqrt{2} G_F m_h^2}, 
\\
 \epsilon^s_{e \tau} & =  \epsilon^{e\mu}_{\mu\tau} & = 
 \frac{f_{\mu\tau} f^*_{e\mu}}{\sqrt{2} G_F m_h^2}, 
\end{eqnarray}
respectively. Notice that  $ \epsilon^m_{\mu\tau} = -  \epsilon^{s*}_{\mu\tau} $, 
since both NSI parameters are related to the couplings $f_{e \mu}$ and $f_{e \tau}$.

As we discuss in section \ref{sec:Numerical-analysis}, the ratios of Yukawa couplings  
$f_{e \mu}/f_{\mu\tau} $ and $f_{e \tau}/f_{\mu\tau} $ are entirely determined by the neutrino 
mixing angles and Dirac phase of the PMNS matrix $U$ -- see eqs. (\ref{fnh}) and (\ref{fih}) --,  so the impact of the 
improved bounds on BR($\mu\rightarrow e \gamma)$ can be easily estimated: 
given that the limit is now $\sim$ 0.05 times smaller than in the study of \cite{Ohlsson:2009vk}, and the contribution of the singly charged scalar $h^+$ to 
BR($\mu\rightarrow e \gamma)$ depends on 
$|f^*_{e \tau}f_{\mu\tau}|^2$, the current constraints on 
$|f_{\alpha \beta}|$ are roughly a factor 2 tighter than before.
Therefore, since the strength of the NSI depends on 
$\epsilon_{\alpha \beta}^{\rho \sigma} \propto f_{\sigma\beta} f^*_{\rho\alpha}$, 
generically we expect that the allowed size of the NSI is reduced by a factor 
$\sim$ 1/4. According to \cite{Ohlsson:2009vk}\footnote{Notice that although the analysis of  \cite{Ohlsson:2009vk} has been done for 
$\kappa =1$, the impact on NSI of the new  bounds from BR($\mu\rightarrow e \gamma)$
(and in general from any LFV decay $\ell_\alpha  \rightarrow \ell_\beta \gamma)$
is independent of the value of $\kappa$ chosen, because they constraint 
directly $|f_{\alpha\sigma}^* f_{\sigma\beta}|/m_h^2$, which is the same combination that 
appears in the NSI parameters, eq.~(\ref{eq:eps}). 
The only effect of increasing  $\kappa$ may be that a given point ($f_{\alpha\sigma},  f_{\sigma\beta}, m_h$) is able to  fit  neutrino masses with  smaller $g_{ab}$ and therefore possibly lighter $m_k$.}, this implies that in the most
favorable case of IH neutrino mass spectrum, $\epsilon^s_{e \tau}$ and 
$\epsilon^s_{\mu\tau}$ are in the range $3 \times(10^{-5} - 10^{-4})$, 
which is in a range difficult to probe, but it might be in a future neutrino factory with a $\nu_\tau$ near 
detector \cite{Tang:2009na}.

\subsection{Bounds on the masses of the charged scalars. }
\label{sec:LHC}

Regarding limits on singly-charged bosons decaying to leptons, the best limit  
still comes from LEP II, $m_h>100$ GeV.

ATLAS and CMS have placed limits on doubly-charged boson masses from searches of dilepton final states, using data samples corresponding to $\sqrt{s} =$ 7 TeV with 
an integrated luminosity  of 4.7  $\text{fb}^{-1}$
and 4.9  $\text{fb}^{-1}$, respectively  \cite{ATLAS:2012hi, Chatrchyan:2012ya}.
The authors of  \cite{delAguila:2013mia}  show that, with current data at $8$ TeV and $20$ $\text{fb}^{-1}$, all the bounds are expected to become about $\sim 100$ GeV more stringent if no significant signal is seen. 
Further tests on the nature of the doubly charged scalar
(i.e., singlet or triplet of $SU(2)_L$) can be obtained by analysing tau lepton decay distributions which are sensitive to the chiral structure of the couplings \cite{Sugiyama:2012yw}. 
The main production mechanisms of doubly-charged bosons at hadron colliders are pair production via an s-channel exchange of a photon or a Z-boson, and associated production with a charged boson via the exchange of a W-boson (see \cite{delAguila:2013yaa,delAguila:2013hla} for a general analysis of the production and detection at LHC of doubly charged scalars belonging to different electroweak representations). In the Zee-Babu model, the associated production is absent, because the new scalars are $SU(2)_L$ singlets. 

The ATLAS analysis \cite{ATLAS:2012hi} focuses on the $ee, \mu\mu, e \mu$ channels and assumes that the rest of the channels can make up to $90\%$ of the total decays. Then, 
the limits for the Zee-Babu model are, at the $95\%$ C.L., $322, 306, 310$ GeV ($151, 176, 151$ GeV) for branching ratios of $100\%$ ($10\%$) to the $ee, \mu\mu, e \mu$ channels.  Notice that in \cite{ATLAS:2012hi} the limits on doubly-charged bosons coupling to left-handed leptons are applied, in addition to the seesaw type II case, to the Zee-Babu model. However, this is not so, as the doubly-charged singlets in the Zee-Babu model are $SU(2)_L$ singlets and thus couple only to right-handed leptons, at variance with the seesaw type II models, where the doubly-charged bosons are $SU(2)_L$ triplets and do couple only to left-handed leptons. 
Therefore, in the  Zee-Babu case they 
have a reduced production cross section,  due to their different couplings to the Z-boson, 
around $2.5$ times smaller than for the case of the triplet \cite{Muhlleitner:2003me}, and less stringent limits apply:  for the Zee-Babu model one should look at the second part of table I of  \cite{ATLAS:2012hi}, the one for $H_R^{\pm\pm}\equiv k^{\pm\pm}$.

The CMS Collaboration has searched for doubly-charged bosons which 
are $SU(2)_L$ triplets, both 
assuming that they decay to the different dilepton final states $\ell \ell$ 
($\ell = e,\mu,\tau$) $100 \%$ of the times, i.e., $\text{BR}(k^{++} \rightarrow \ell\ell)=1$,
and also considering  several benchmark points with different branching ratios.

The CMS  $95\%$ C.L. limits for pair production of $SU(2)_L$ singlets, which is the one relevant for the Zee-Babu Model, are around $60-80$ GeV less stringent \cite{Muhlleitner:2003me, delAguila:2013hla}:

\begin{itemize}
\item $ee, \mu\mu, e \mu$ :           $~310$ GeV,
\item $e\tau, \mu\tau$ :           $~220$ GeV,
\item $\tau \tau$ :           $~100$ GeV.
\end{itemize}

Note that whenever the branching ratio to $\tau \tau$ is less than $30\%$ (see table I and VI of \cite{Chatrchyan:2012ya}), the bounds are $\sim280$ GeV, provided that there is a significant fraction of decays into light leptons ($ee, \mu\mu, e \mu$). 

In the Zee-Babu model the decay width of $k^{\pm\pm}$ into same sign leptons is given by
\begin{equation}
\Gamma(k \rightarrow \ell_a \ell_b) = \frac{|g_{ab}|^2}{4 \pi (1+\delta_{ab})} m_k \ . 
\end{equation}
Since the $g_{ab}$ couplings are free parameters, the BRs of the different decay modes are 
a priori unknown, so we can not apply directly these bounds. 
As we will see in the numerical analysis, section \ref{sec:Numerical-analysis}, 
once neutrino oscillation  data and low energy constraints are taken into account, 
the branching ratio to $\tau \tau$ is very small in the Zee-Babu model, less than about $1\%$. Then, a conservative limit is $m_k>220$ GeV.  

Moreover, in the ZB model for $m_k>2\, m_h>200$ GeV, it can happen that the doubly charged scalar decays predominantly into $h h$, which can easily escape detection. 
This way the constraints from dilepton searches could be evaded. 
The relevant decay width is given by
\begin{equation}
\Gamma(k \rightarrow h h ) = \frac{1}{8 \pi} 
\left[ \frac{\mu}{m_k} \right]^2  m_k \, \sqrt{1- \frac{4 m_h^2}{m_k^2}} \ .
\end{equation}
Then,   
even  for $g_{ab} \sim 1$,  
for $m_h=100$ GeV and $m_k=200$ GeV, we have that $\frac{\Gamma(k\rightarrow hh)}{\Gamma(k\rightarrow \ell\ell)} \geq 1$ for $\mu\geq m_k$, which is still natural as long it is not very large. Thus, we take  $m_k \geq 200$ GeV in the numerical analysis.

\subsection{Stability of the potential. \label{sub:Stability}}

In this section we consider further constraints on the ZB model parameter space coming from vacuum stability conditions. The Hamiltonian in quantum mechanics has to be bounded from below, this requires that the quartic part of the scalar potential in eq.~(\ref{eq:V}) should be positive for all values of the fields and for all scales. Then, if two of the fields $H,k$ or $h$ vanish one immediately finds\footnote{We do not consider the possibility of zero couplings, which can only appear at very specific scales.}:
\begin{equation}
\label{eq:s1}
 \lambda_H > 0, \qquad \lambda_{h} > 0, \qquad  \lambda_{k} > 0  \, .
\end{equation}
Moreover the positivity of the potential whenever one of the scalar fields $H,h,k$ is zero implies  
\begin{equation}
\label{eq:s2}
\alpha,\beta,\gamma > -1\,,
\end{equation}
where we have defined
\begin{equation}
\label{eq:abg-defs}
\alpha = \lambda_{hH}/(2 \sqrt{ \lambda_H \lambda_{h}})\,,\,
\beta = \lambda_{kH}/(2 \sqrt{ \lambda_H \lambda_{k}})\,,\, 
\gamma = \lambda_{hk}/(2 \sqrt{ \lambda_h \lambda_{k}})\,.
\end{equation}
Eq.~\eqref{eq:s2} constrains only negative mixed couplings, 
$\lambda_{xH}, \lambda_{hk}$ ($x=h,k$), since for positive ones  the potential is definite positive
and only the perturbativity limit, $\lambda_{xH},\lambda_{hk} \lesssim 4 \pi$ applies.
Finally, if at least two of the mixed couplings are negative, 
there is an extra constraint, which  can be written as:
\begin{equation}
\label{eq:s3} 
1-\alpha^2 -\beta^2 - \gamma^2 + 2 \alpha \beta \gamma > 0 
\qquad \vee \qquad \alpha+\beta+\gamma > -1\, . 
\end{equation} 
We have checked that the above conditions, eqs.~(\ref{eq:s1}, \ref{eq:s2}, \ref{eq:s3}), are equivalent to the ones derived in 
\cite{Kanemura:2000bq} for the Zee model, but they differ from the ones used in 
\cite{Chao:2012xt} for the ZB model, which seem not to be symmetric under 
the exchange of $\alpha, \beta, \gamma$, as they should.
Our constraints also agree with the results obtained by using copositive criteria (see for instance \cite{Kannike:2012pe}).

The discovery of the Higgs boson with mass $m_H\sim$ 125 GeV at the LHC has raised the interest on the vacuum stability of the SM potential: for the current central values of the strong coupling constant and the Higgs and top quark masses, the 
Higgs self-coupling $\lambda_H$ would turn negative at a scale $\Lambda \sim 10^{10}-10^{13}$ 
GeV \cite{Degrassi:2012ry}, indicating the existence of new physics beyond the SM below that scale. In fact, by using
state of the art radiative corrections, the authors of \cite{Degrassi:2012ry} find that absolute stability of the SM Higgs potential up to the Planck scale 
is excluded at 98\% C.L. for $m_H < 126$~GeV.  

\begin{figure}
	\centering
	\includegraphics[width=0.45\textwidth]{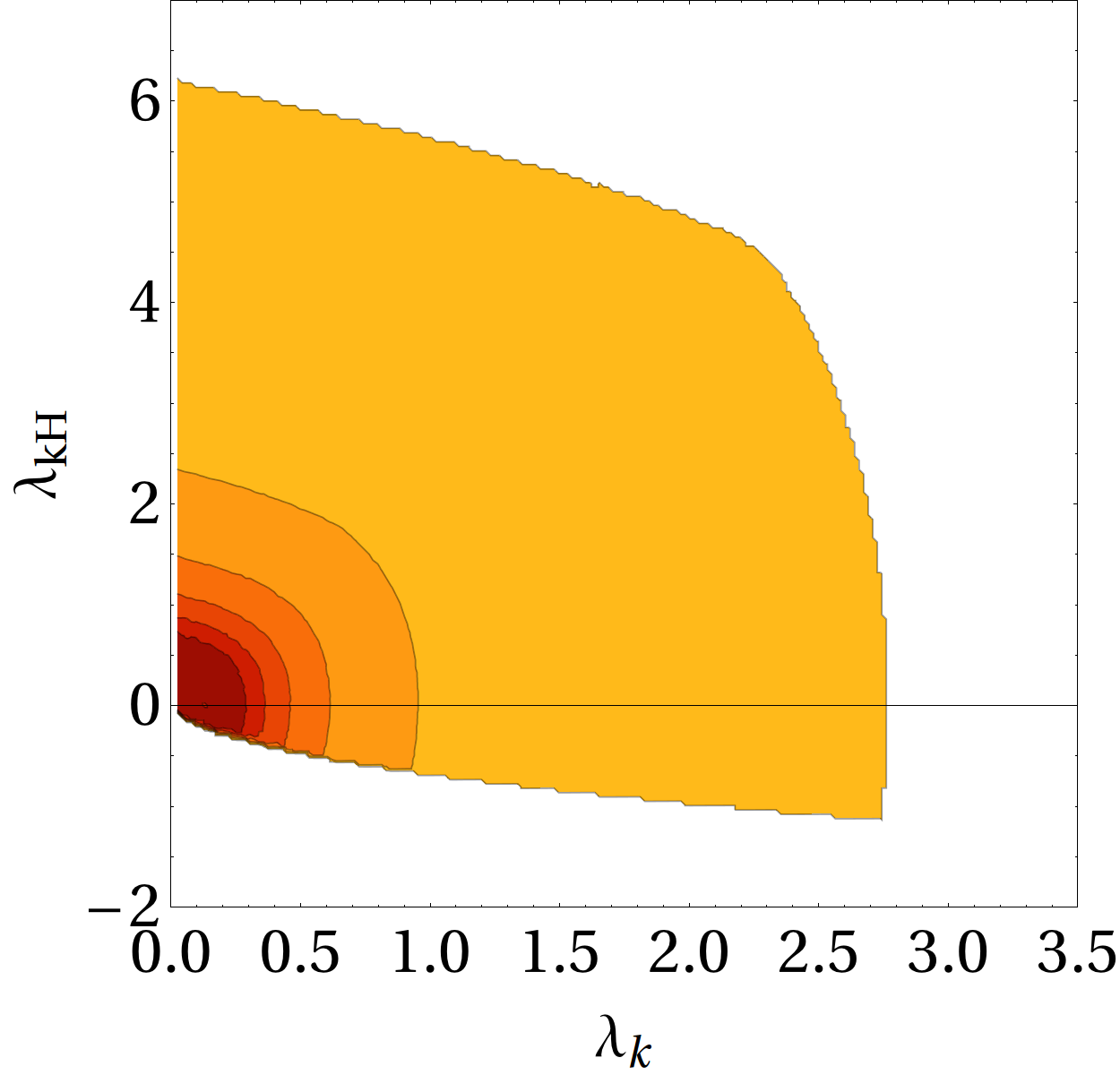}~~
	\includegraphics[width=0.45\textwidth]{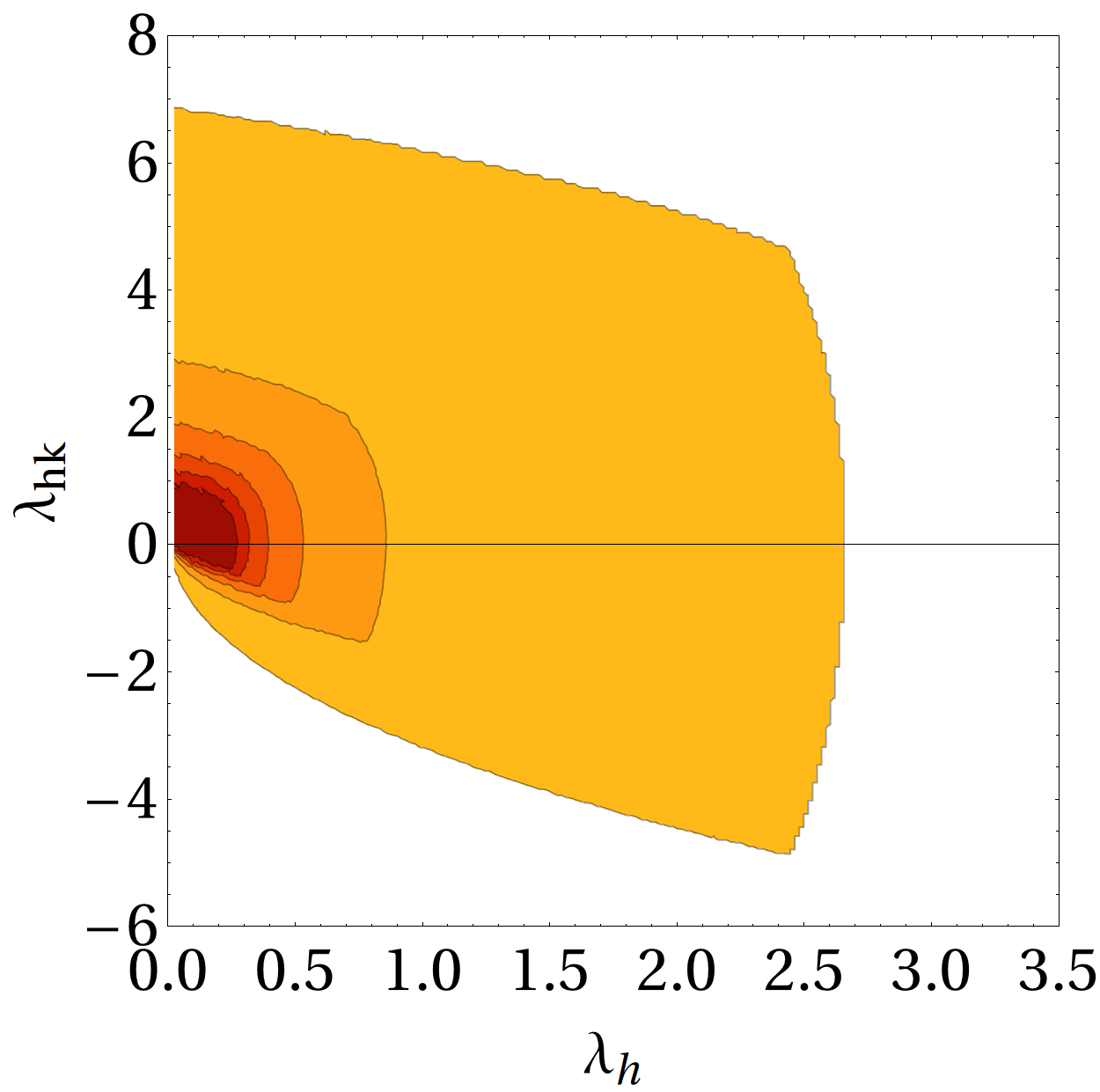}
	\caption{Allowed regions in $\lambda_{kH}$ vs $\lambda_k$ (left) and $\lambda_{hk}$ vs $\lambda_h$ (right), taken at the $m_Z$ scale, if perturbativity/stability is required to be valid up to $10^3,10^6,10^9,10^{12},10^{15},10^{18}$ GeV (from light to dark colours). The rest of the parameters entering the RGE are taken at their measured value or varied in the range allowed by the perturbativity/stability requirement up to the given scale.} \label{fig:perturbativity}
\end{figure}

The  one-loop renormalization group equations  (RGEs) in the ZB model are written in Appendix \ref{RGE}. For a given set of parameters defined at the electroweak scale, 
and satisfying the stability conditions discussed above, 
we calculate the running couplings numerically by using one-loop RGEs.
From eqs.~(\ref{eq:RGES}), we see that the new scalar couplings $\lambda_{hH},\lambda_{kH}$ 
always contribute positively to the running of the Higgs quartic coupling 
$\lambda_H$, compensating for the large and negative contribution of the top quark Yukawa coupling. 
Therefore, the vacuum stability problem can be alleviated in the ZB model with $\lambda_H$ remaining positive up to the Planck scale for the present central values of $m_t$ and $m_H$ if $\lambda_{xH}$ are not extremely small ($\lambda_{xH}\sim \pm 0.2$ are enough to stabilize $\lambda_H$ maintaining stability/perturbativity of all couplings up to the Planck scale (see fig.~\ref{fig:perturbativity})).

On the other hand, as we discuss in section \ref{sec:gamma}, the slight  excess in the 
Higgs diphoton decay channel found at LHC can be accommodated in the ZB model with relatively light singlet scalars and large, negative, mixed couplings 
$\lambda_{hH},\lambda_{kH}$. However for  such values of the scalar couplings at the 
electroweak scale, the RGEs lead to vacuum instability 
($2 \sqrt{ \lambda_H \lambda_{x}} + \lambda_{xH} < 0$, $x=h,k$) and/or non-perturbativity  
($\lambda_x > 4 \pi$) well below the Planck scale. This can be seen in fig.~\ref{fig:perturbativity} where we have performed a complete scan of the quartic couplings of the scalar potential, run all of them from $m_Z$ up to a given scale ($\mu=10^{3n}$ GeV with $n=1,2,\cdots,6$), and check that stability (as explained before) and perturbativity ($\lambda_i < 4\pi$) are satisfied at all scales below $\mu$. On the left we represent the region allowed in the $\lambda_{kH}$--$\lambda_{k}$ plane, with $\lambda$'s taken at the $m_Z$ scale, when stability/perturbativity is imposed up to the different scales $\mu$. Lighter regions correspond to small scales and obviously include the regions of larger scales. A similar plot is obtained for $\lambda_{hH}$ vs $\lambda_{h}$. On the right we present the equivalent results for the couplings $\lambda_{hk}$ vs $\lambda_{h}$.

\section{$H \rightarrow \gamma \gamma$ and $H \rightarrow Z \gamma$ 
\label{sec:gamma}}

It remains an open question whether the $125$ GeV Higgs boson discovered by ATLAS \cite{Aad:2012tfa}  and CMS \cite{Chatrchyan:2012ufa} is the SM one or has some extra features coming from new physics. While all the present measurements of the Higgs properties are consistent with the SM values, the uncertainties are still large, so there is plenty 
of room for non-standard signals to show up in the upcoming 13-14 TeV run data.
Moreover, the present experimental  situation of the 
$H \rightarrow \gamma\gamma$ decay channel is far from clear: although the 
last reported analysis of the CMS and ATLAS Collaborations on the diphoton 
signal strength are barely consistent with each other within $2\sigma$, 
 ATLAS still observes a $\sim 2\sigma$ excess over the SM prediction \cite{ATLAS:2013oma},
while the CMS measurement has become consistent with the SM at $1 \sigma$ \cite{CMS:ril}:
\begin{eqnarray} \label{ggdata}
{\rm ATLAS:} & R_{\gamma\gamma} = 1.55^{+0.33}_{-0.28}\,, &
\nonumber \\
{\rm CMS:} & R_{\gamma\gamma} = 0.78^{+0.28}_{-0.26}\,,  &{\rm MVA \ analysis}
\\
{\rm CMS:} & R_{\gamma\gamma} = 1.11^{+0.32}_{-0.31}\,.   & {\rm cut \  based \  analysis}
\nonumber
\end{eqnarray}

It is thus worthwhile to explore whether  an eventually confirmed deviation from the SM
prediction in the $H \rightarrow \gamma \gamma$ channel can be accommodated within the ZB model.

In the SM the $H \rightarrow \gamma \gamma$ channel is dominated by the $W$ boson loop 
contribution, which interferes destructively with the top quark one. Since the Higgs coupling to photons is induced at the loop-level, 
extra charged fermions or scalars with significant couplings to the Higgs can change drastically the  $H \rightarrow \gamma \gamma$ channel with respect to the Standard Model expectations, either enhancing it or reducing it \cite{Carena:2012xa}. Moreover, in the absence of direct signatures of new particles at LHC, the enhanced Higgs diphoton decay rate might provide
an indirect hint of physics beyond the SM. 

The value of the $H \rightarrow \gamma \gamma$ decay width in the ZB model with respect to the SM one is given by \cite{Ellis:1975ap,Shifman:1979eb,Carena:2012xa}:
\begin{equation}
R_{\gamma \gamma} = \frac{\Gamma(H \rightarrow \gamma \gamma)_{ZB}}{\Gamma(H \rightarrow \gamma \gamma)_{SM}}= \left | 1+ \delta R (m_h,\lambda_{hH})+4 \, \delta R (m_k,\lambda_{kH}) \right |^2\,, \label{eq:ggratio}
\end{equation}
where we have defined $\delta R(m_x,\lambda_{xH})$ for the scalar $x$ with mass $m_x$ and coupling to the Higgs $\lambda_{xH}$ as:
\begin{equation}
\delta R (m_x,\lambda_{xH}) \equiv \frac{\lambda_{xH} \, v^2}{2  m_x^2}\frac{A_0(\tau_x)}{A_1(\tau_W)+\frac{4}{3} A_{1/2}(\tau_t)}\,, \label{eq:deltaR}
\end{equation}
with $\tau_i\equiv \frac{4 m_i^2}{m_H^2}$ and the loop functions $A_i(x)$ $(i=0,1/2,1)$ are 
defined in Appendix \ref{loop}. 
Notice that the dominant $W$ contribution is $A_1(\tau_W) = - 8.32$ for a Higgs mass 
of 125 GeV, while $A_0(\tau_{h,k}) > 0$, 
therefore in order to obtain a constructive interference we need to consider 
negative couplings  $\lambda_{hH}, \lambda_{kH}$. 

As discussed in sec. \ref{sub:Stability}, stability of the potential imposes  that 
$2 \sqrt{ \lambda_H \lambda_{x}} + \lambda_{xH} > 0$, for $x=h,k$. Since  
$M_H \sim$ 125  GeV fixes the value of the Higgs self-coupling to $\lambda_H \sim 0.13$, 
it is immediately apparent that large and negative $\lambda_{xH}$ couplings 
are going to be in conflict with stability of the potential, unless we push 
$\lambda_x$ close to the naive perturbative limit ($\lambda_x < 4 \pi$), 
 for which  $-3 \lesssim \lambda_{hH},\lambda_{kH}$. 
Notice that this fact is not a special feature of the ZB model, but a generic problem of any 
scenario in which the enhancement of the Higgs diphoton decay rate is 
due to a virtual charged scalar.

We can consider three different cases: 
\begin{itemize}
\item {If $m_h \ll m_k$,
\begin{equation}
R_{\gamma \gamma}^h \approx  \left |1+\delta R (m_h,\lambda_{hH}) \right |^2 \,;
\label{eq:ggratioa}
\end{equation}
}
\item {If $m_k \ll m_h$,
\begin{equation}
R_{\gamma \gamma}^k  \approx  
\left |1+ 4  \delta R (m_k, \lambda_{kH})\right |^2  \,;
\label{eq:ggratiob}
\end{equation}
}
\item {If $m_h \approx m_k \equiv m_S$, with 
\begin{equation}
R_{\gamma \gamma}^S  \approx  
\left | 1+ \delta R (m_S,\lambda_{hH})+4 \, \delta R (m_S,\lambda_{kH}) \right |^2\,.
\end{equation}
}
\end{itemize}

For the same masses and couplings of both singlets, the doubly charged produces a larger enhancement/suppression than the singly-charged, due to its greater charge.

\begin{figure}
	\centering
	\includegraphics[width=0.6\textwidth]{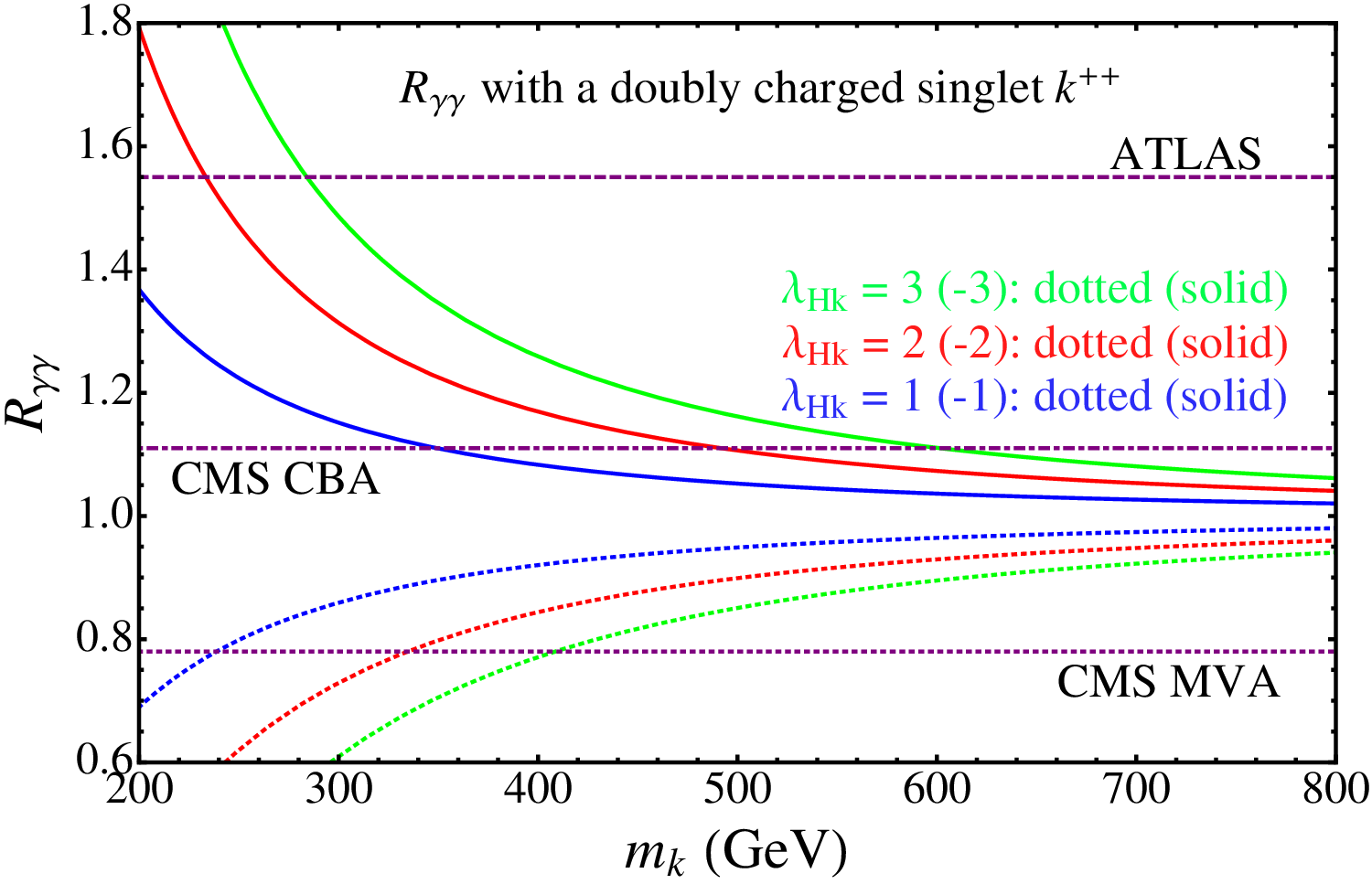}
	\caption{$R_{\gamma \gamma}$ in the presence of a doubly charged particle. Both an enhancement (as seen by ATLAS \cite{ATLAS:2013oma}) or a suppression (as seen by CMS \cite{CMS:ril}), can be accommodated. For the same masses and couplings, the singly-charged produces a smaller enhancement/suppression than the doubly-charged, due to its smaller charge.} \label{doubly}
\end{figure}

\begin{figure}
	\centering
        \includegraphics[width=0.45\textwidth]{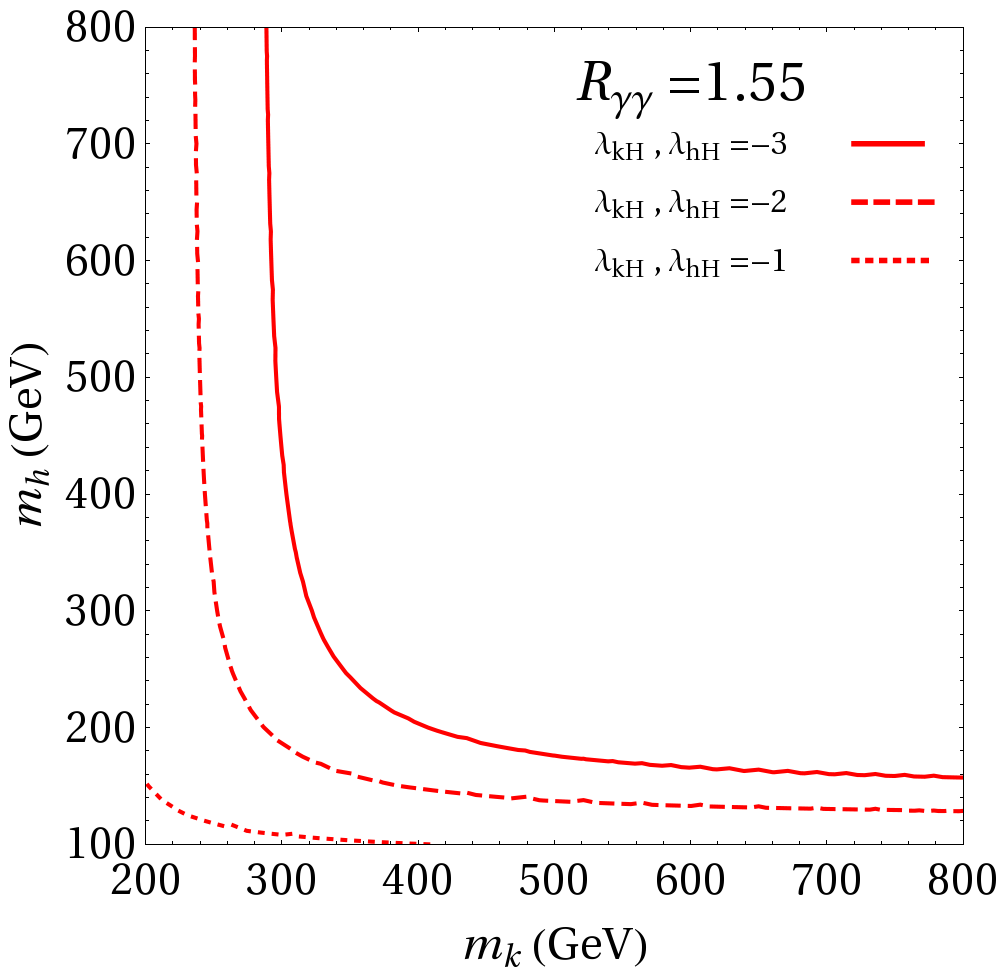}~~
	\includegraphics[width=0.45\textwidth]{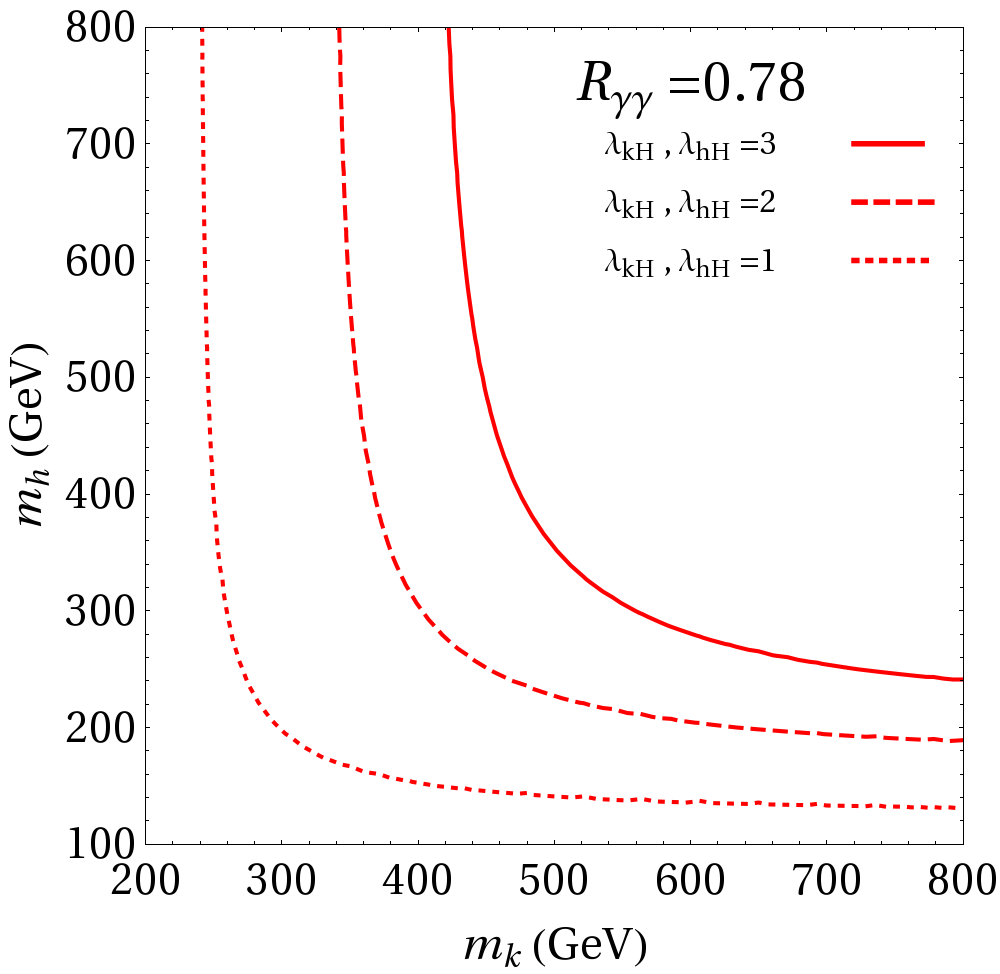}
	\caption{Contour of $R_{\gamma\gamma}=1.55$ (left) \cite{ATLAS:2013oma} and $R_{\gamma\gamma}=0.78$ (right) \cite{CMS:ril} in the presence of a singly charged and doubly charged particle with the same couplings.} \label{contourgg}
\end{figure}

The largest enhancement can happen when both charged scalars are about the same mass and these masses are low enough. 
We show in fig.~\ref{doubly} the prediction of the ratio 
$R_{\gamma \gamma}$ when the doubly charged scalar $k$ dominates, 
for different values of the coupling with the Higgs, $\lambda_{kH}$. Both an enhancement (as seen by ATLAS \cite{ATLAS:2013oma}) or a suppression (as seen by CMS \cite{CMS:ril}), can be accommodated. In fact, deviations from the SM value are expected, i.e., $R_{\gamma \gamma}\neq 1$, in particular for below the TeV scale singlets and sizeable scalar couplings. Of course, even for light singlets it is possible that $R_{\gamma \gamma}\approx 1$, either because the relevant scalar couplings are tiny or due to
a cancellation between the contributions of the singly charged and the doubly charged scalars.

In principle, the enhancement $R_{\gamma \gamma}$ 
induced by a singly charged scalar $h$ of similar mass and coupling to the Higgs
$\lambda_{hH} \sim \lambda_{kH}$ is smaller; however 
since the lower limit on $m_h$ from  LEP II direct searches is weaker
$m_h > 100$ GeV, as discussed in the previous section, 
and the largest contribution occurs for lower masses, the resulting values 
of $R_{\gamma \gamma}$ for the allowed range of $m_h$ are 
comparable to the doubly charged case.

We show in fig.~\ref{contourgg} the contours of  $R_{\gamma\gamma}=1.55$ ($0.78$), motivated by the experimental results of ATLAS and CMS \cite{ATLAS:2013oma,CMS:ril}, in the plane of the singly and doubly charged masses, for various negative (positive) couplings. 
In summary, 
to obtain $R_{\gamma \gamma} \sim 1.5$ we need $m_h \lesssim 200 \, \textrm{GeV}$ and/or
$m_k \lesssim 300  \, \textrm{GeV}$.
As it will be shown in the numerical analysis section, these scalar masses are in
tension with describing  neutrino oscillation data  and being compatible with current low-energy bounds in the ZB model if naturality is required at the level of $\kappa=1$, especially for the NH spectrum.  
Moreover, the large negative values of the couplings $\lambda_{xH} \sim -2$ required to obtain such enhancement
also induce vacuum instability of the ZB scalar potential, unless  
the corresponding coupling $\lambda_{x}$ is close to the perturbative limit, $\lambda_{x} \sim 8$. 

\begin{figure}
	\centering
	\includegraphics[width=0.6\textwidth]{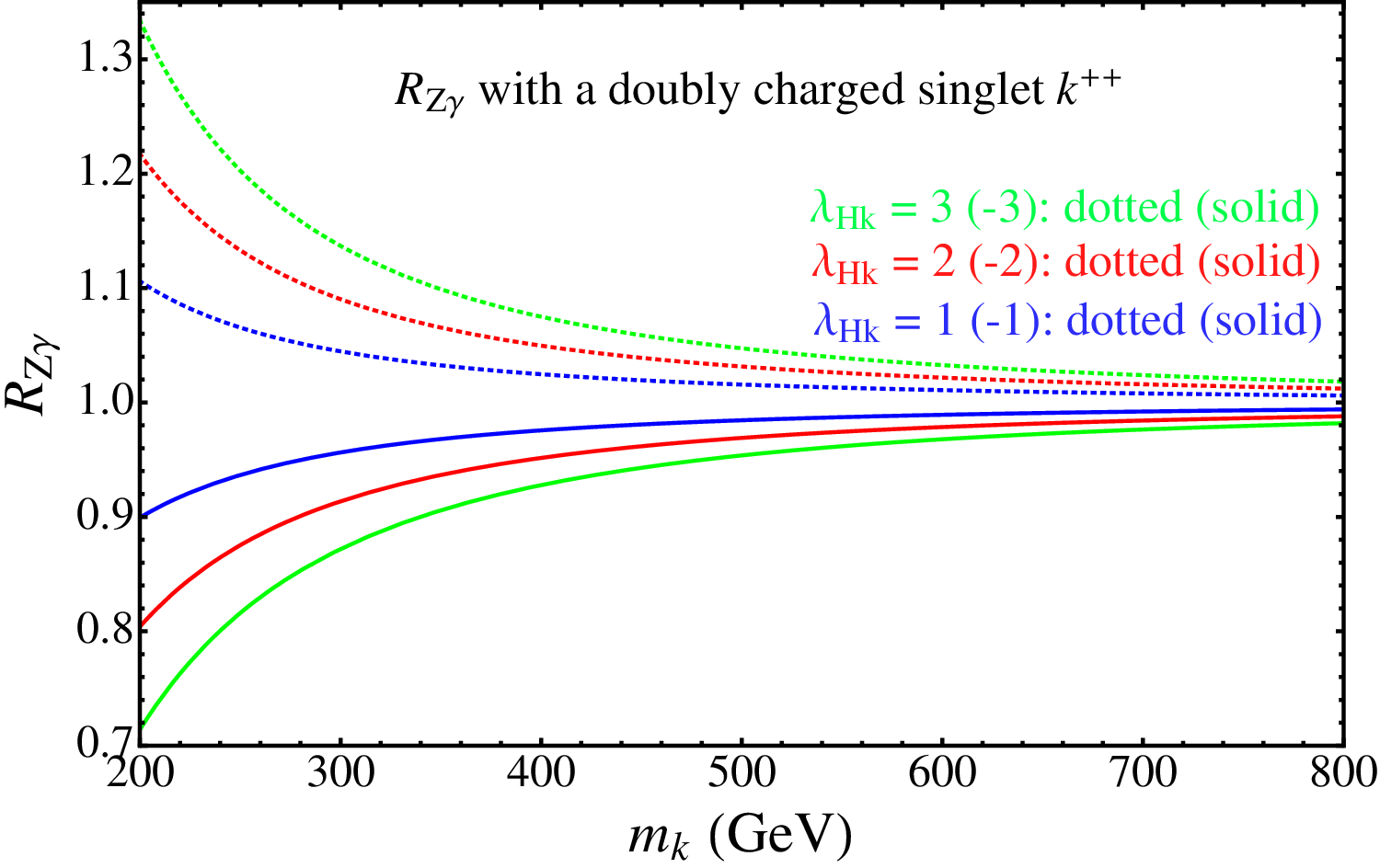}
	\caption{$R_{\gamma Z}$ in the presence of a doubly charged particle. As can be seen, $H\rightarrow Z \gamma$ is anticorrelated with respect to $H\rightarrow \gamma \gamma$.} \label{Zg}
\end{figure}

There is a correlation between $H \rightarrow \gamma \gamma$ and $H \rightarrow Z \gamma$
\cite{Gunion:1989we, Carena:2012xa, Chen:2013vi}.
The ratio of the $H \rightarrow Z \gamma$ decay rate in the ZB model with respect to the SM 
one is:
\begin{equation}
R_{Z \gamma} = \frac{\Gamma(H \rightarrow Z \gamma)_{ZB}}{\Gamma(H \rightarrow Z  \gamma)_{SM}}= \left | 1-  g_{Zhh}  \frac{\lambda_{hH} \, v^2}{m_h^2}
\frac{A_0(\tau_h,\lambda_h)}{\mathcal{A}_{SM}^{Z\gamma}}
-   g_{Zkk} \frac{2 \lambda_{kH} \, v^2}{m_k^2} 
\frac{A_0(\tau_k,\lambda_k)}{\mathcal{A}_{SM}^{Z\gamma}} \,,
 \right |^2 \label{eq:zgratio}
\end{equation}
where $\mathcal{A}_{SM}^{Z\gamma}$  is the SM $H \rightarrow Z \gamma$ decay 
amplitude, 
\begin{equation}
\mathcal{A}_{SM}^{Z\gamma} = 
\cot \theta_W A_1(\tau_W,\lambda_W) 
+ 6 Q_t \frac{T_3^t  - 2 Q_t s^2_W}{s_W c_W} A_{1/2}(\tau_t,\lambda_t) \,,
\end{equation}
with $\lambda_i\equiv \frac{4 m_i^2}{m_Z^2}$,  
and the $Z$ boson couplings to the new charged scalars are 
$g_{Zxx} = - Q_x \cot\theta_W$, $x=h,k$.
The loop functions $A_i(x,y)$ $(i=0,1/2,1)$ can be found  in Appendix \ref{loop}.

In fact, to have an enhancement in the $H \rightarrow \gamma \gamma$ channel, we need negative couplings of the singlets with the Higgs, which in turn implies that the $H \rightarrow Z \gamma$ channel is reduced with respect to SM prediction, as can be seen in  fig. \ref{Zg}.

\section{Analytical estimates}
\label{sec:ae}

In this section we give some order of magnitude estimates of the free parameters in the ZB model, 
which complement and help to understand our full numerical analysis.  
In particular,  we want to estimate to which extent light charged scalar masses,  
for instance like those required to fit an enhanced Higgs diphoton decay rate or to have a chance of being discovered at the LHC, are consistent with neutrino oscillation data and low-energy constraints.

As discussed in sec. \ref{sec:ZB-model}, with respect to the SM the ZB model has  17 extra  parameters relevant for neutrino masses (9 moduli and 5 phases from the Yukawa couplings 
$f,g$, and 3 mass parameters from the charged scalar sector, $m_h, m_k$ and $\mu$), plus 5 quartic couplings in the scalar potential. However, some of the free parameters can be 
traded by the measured neutrino masses and mixings, 
ensuring  in this way that the experimental data is reproduced and reducing the number of free variables as follows. 

Since $\det f = 0$, there is an eigenvector ${\bf a} = (f_{\mu\tau}, - f_{e \tau},f_{e \mu})$ 
which corresponds to the zero eigenvalue,  $f {\bf a} =0$  \cite{Babu:2002uu}. 
Then,  by exploiting the fact that {\bf a} is also an eigenvector of $\mathcal{M}_{\nu}$, 
we have 
\begin{equation}
D_\nu U^T {\bf a} = 0, 
\end{equation}
which leads to  three equations, one of which is trivially satisfied because one element 
of $D_\nu$ is zero.
The other two equations allow to write  the ratios of  Yukawa couplings $f_{ij}$ in terms of the 
neutrino mixing angles and Dirac phase as follows:
\begin{eqnarray} 
\frac{f_{e \tau}}{f_{\mu\tau}} &=& \tan \theta_{12} \frac{\cos \theta_{23}}{\cos\theta_{13}} + 
\tan \theta_{13} \sin \theta_{23} e^{-i \delta}\,,
\nonumber \\
\frac{f_{e \mu}}{f_{\mu\tau}} &=& \tan \theta_{12} \frac{\sin \theta_{23}}{\cos\theta_{13}} -
\tan \theta_{13} \cos \theta_{23} e^{-i \delta}\,,
\label{fnh}
\end{eqnarray}
in the NH case,  and 
\begin{eqnarray} 
\frac{f_{e \tau}}{f_{\mu\tau}} &=& -\frac{\sin \theta_{23}}{ \tan \theta_{13}} e^{-i \delta}\,,
\nonumber \\
\frac{f_{e \mu}}{f_{\mu\tau}} &=& \frac{\cos \theta_{23}}{\tan\theta_{13}}  e^{-i \delta}\,,
\label{fih}
\end{eqnarray}
for IH spectrum. Therefore, we choose $f_{\mu\tau}$ as a free, real,  parameter and 
obtain (complex) $f_{e\mu}$ and $f_{e\tau}$ from the above equations. Notice 
that the measured values, $s_{12}^2 \sim 0.3$, $s_{23}^2 \sim 0.4$ and 
$s_{13}^2 \sim 0.02$ imply that, for NH, the first term on the right-hand side of 
eqs.~(\ref{fnh}) dominates  and leads to 
$f_{e\mu} \sim  f_{\mu\tau}/2  \sim f_{e\tau}$. 
Conversely, for IH it is clear that  
$f_{e\tau}/ f_{e\mu} =  - \tan\theta_{23} \sim -1$
and $|f_{e\mu}/ f_{\mu\tau}| \sim |f_{e\tau}/ f_{\mu\tau}| \sim 4$.
Of course, to explain such fine-tuned relations of Yukawa couplings a complete 
theory of flavour would be needed, which is beyond the scope of this work. 

Regarding the Yukawa couplings $g$, we keep $\gee, \gem$ and $\get$ as free 
complex parameters and fix the remaining ones ($g_{\mu\mu},g_{\mu\tau},g_{\tau\tau}$) 
by imposing the equality of the three elements $m_{22}, m_{23}$
and $m_{33}$ of the neutrino mass matrix $\mathcal{M}_{\nu}$,  written in terms of the 
parameters of the ZB model in eq.~(\ref{eq:MnuYukawas}), and in terms of the masses and mixings 
measured in neutrino oscillation experiments in eq.~(\ref{eq:numass}),  i.e.,  
\begin{equation}
m_{ij}=(U D_\nu U^T)_{ij} = \zeta f_{ia} \omega_{ab} f_{jb}, 
\label{eq:mij}
\end{equation}
where we have defined $\omega_{ab} \equiv m_a g^*_{ab} m_b$, 
and $\zeta = \frac{\mu}{48 \pi^2 M^2} \tilde{I}(r)$,  being $r$
the ratio of the scalar masses, $r\equiv m_{k}^{2}/m_{h}^{2}$. 

Because of the hierarchy among the charged lepton masses, $m_e \ll m_\mu, m_\tau$, 
it is natural to assume that $\omega_{ee}, \omega_{e\mu},\omega_{e\tau} 
\ll \omega_{\mu\mu}, \omega_{\mu\tau}, \omega_{\tau\tau}$. 
Within the approximation $\omega_{ea} =0$, the equation (\ref{eq:mij})  for neutrino masses  is simplified, and we can easily estimate the ranges of parameters consistent with 
neutrino oscillation data. Thus in this section
we neglect them, although we keep all $\omega_{ab}$ in the full numerical analysis\footnote{We find that, in general, this is a very good approximation.} 
We then have
\begin{equation}
m_{22}\simeq\zeta f_{\mu\tau}^{2}\omega_{\tau\tau},~m_{23}\simeq-\zeta f_{\mu\tau}^{2}\omega_{\mu\tau},~m_{33}\simeq\zeta f_{\mu\tau}^{2}\omega_{\mu\mu}.
\label{eq:mij_app}\end{equation}
From the large atmospheric angle we expect  
\begin{equation}
\label{eq:omegas}
|\omega_{\tau\tau}|\simeq|\omega_{\mu\tau}|\simeq|\omega_{\mu\mu}|,
\end{equation}
which leads to a definite hierarchy among the corresponding $g_{ab}$ couplings: 
\begin{equation}
g_{\tau\tau}:g_{\mu\tau}:g_{\mu\mu}\sim m_{\mu}^{2}/m_{\tau}^{2}:m_{\mu}/m_{\tau}:1.
\label{couplinghierarchy}\end{equation}

It is now convenient to write the  mass matrix elements $m_{ij}$ in terms of the 
neutrino masses and mixings. 
In the normal hierarchy case this gives 
\begin{eqnarray}
\zeta f_{\mu\tau}^{2}\omega_{\tau\tau} & \simeq & m_{3}c_{13}^{2}s_{23}^{2}+m_{2}e^{i\phi}(c_{12}c_{23}-e^{i\delta}s_{12}s_{13}s_{23})^{2}\:,\nonumber \\
\zeta f_{\mu\tau}^{2}\omega_{\mu\tau} & \simeq & -m_{3}c_{13}^{2}c_{23}s_{23}+m_{2}e^{i\phi}(c_{12}s_{23}+e^{i\delta}c_{23}s_{12}s_{13})(c_{12}c_{23}-e^{i\delta}s_{12}s_{13}s_{23})\:,\nonumber \\
\zeta f_{\mu\tau}^{2}\omega_{\mu\mu} & \simeq & m_{3}c_{13}^{2}c_{23}^{2}+m_{2}e^{i\phi}(c_{12}s_{23}+e^{i\delta}c_{23}s_{12}s_{13})^{2} \ , 
\label{eq:mijNangulos}\end{eqnarray}
which for  $m_{3}\simeq0.05$ eV and $m_{2}\simeq0.009$ eV, 
leads to 
\begin{equation}
\zeta f_{\mu\tau}^{2} |\omega_{ab}|  \simeq 0.025 \ \mathrm{eV} \, , \qquad a,b=\mu,\tau,
\label{eq:omegasNH}\end{equation}
in agreement with the expectations of eq.~(\ref{eq:omegas}).

In the inverted hierarchy case, eqs. (\ref{eq:mij_app}) read 
\begin{eqnarray}
\zeta f_{\mu\tau}^{2}\omega_{\tau\tau} & \simeq & m_{1}(c_{23}s_{12}+e^{i\delta}c_{12}s_{13}s_{23})^{2}+m_{2}e^{i\phi}(c_{12}c_{23}-e^{i\delta}s_{12}s_{13}s_{23})^{2}\,,\nonumber \\
\zeta f_{\mu\tau}^{2}\omega_{\mu\tau} & \simeq & m_{1}(s_{12}s_{23}-e^{i\delta}c_{12}c_{23}s_{13})(c_{23}s_{12}+e^{i\delta}c_{12}s_{13}s_{23})\nonumber \\
 & + & m_{2}e^{i\phi}(c_{12}s_{23}+e^{i\delta}c_{23}s_{12}s_{13})(c_{12}c_{23}-e^{i\delta}s_{12}s_{13}s_{23})\,,\label{eq:mijIangulos}\\
\zeta f_{\mu\tau}^{2}\omega_{\mu\mu} & \simeq & m_{1}(s_{12}s_{23}-e^{i\delta}c_{12}c_{23}s_{13})^{2}+m_{2}e^{i\phi}(c_{12}s_{23}+e^{i\delta}c_{23}s_{12}s_{13})^{2},
\nonumber \end{eqnarray}
where $m_{1}\simeq m_{2}\simeq0.05$ eV. It is important to notice that  
for $e^{i\phi} \sim e^{i\delta}  \sim 1$ the matrix elements $m_{ij}$ are of the same order 
as in the NH spectrum,  i.e., 
\begin{equation}
\zeta f_{\mu\tau}^{2} |\omega_{ab}|\simeq 0.025 \:\text{eV}\, ,  \qquad a,b=\mu,\tau.
\label{eq:omegasIH1}\end{equation}
and therefore the hierarchy of couplings  in eq.~(\ref{couplinghierarchy})
is also obtained. 
However, in the IH case  there is a strong cancellation for Majorana 
phases close to $\pi$, so we can obtain smaller values of $\omega_{ab}$. In particular, 
for $\phi=\delta=\pi$ and the best fit values of the masses and mixing angles we find 
\begin{equation}
\zeta f_{\mu\tau}^{2} |\omega_{\mu\mu}|\simeq 0.003  \:\text{eV},
\label{eq:omegasIH}\end{equation}
which allows for a smaller $g_{\mu\mu}$ and, as a consequence, a lighter $m_k$ still 
consistent with the experimental limits. On the contrary, if $\phi\sim \pi$ and $\delta \sim 0$, 
$|\omega_{\tau\tau}|$ can be very small and therefore 
$g_{\tau\tau} \ll (m_{\mu}^{2}/m_{\tau})^{2} \, g_{\mu\mu}$, 
although this cancellation has no phenomenological impact. 
Therefore, although in the following analytic approximations we assume the hierarchy of couplings in eq.~(\ref{couplinghierarchy}),  one has to keep in mind that a larger parameter 
space is expected to be allowed when 
$\phi \simeq \delta \simeq \pi$. Indeed we will confirm in the full numerical analysis
of section \ref{sec:Numerical-analysis}
that this region is specially favoured for light $m_k$.

Now we can estimate the lowest scalar masses able to reproduce current neutrino data. 
Using the neutrino mass equation  we can write\footnote{Notice that similar limits are derived from any of the 23 block elements 
of $\mathcal{M}_{\nu}$ when assuming the hierarchy of the $g$ couplings given in 
eq.~(\ref{couplinghierarchy}).}
\begin{equation} 
\label{eq:numassapp}
\frac{m_{33}}{0.05 \, \text{eV}} \simeq  500   |g_{\mu\mu}| |f_{\mu\tau}|^2  \frac{\mu}{M} 
\frac{\text{TeV}}{M} \tilde{I}(r).
\end{equation}
The upper bound on  $\tau \rightarrow 3\mu$ decay 
implies that 
$|g_{\mu\mu}| \lesssim 0.4 (m_k/ \text{TeV})$, while  
the new  MEG limits on $\mu\rightarrow e \gamma$ lead to 
$\epsilon |f_{\mu\tau} |^2  \lesssim   1.3 \cdot 10^{-3}  (m_h/ \text{TeV})^2$, 
where  $\epsilon \equiv |f_{e \tau}/ f_{\mu\tau}| \sim$  1/2 (4) for NH (IH). 
Substituting these constraints in eq.~(\ref{eq:numassapp}) we obtain 
\begin{equation} 
\label{eq:numasslimit}
\frac{m_{33}}{0.05 \, \text{eV}} \lesssim  0.26 \,  \frac{\mu\, m_k } {\epsilon \, M^2} 
\left(\frac{m_h}{\text{TeV}}\right)^2  \tilde{I}(r), 
\end{equation}
which can be translated into a lower bound on the scalar masses. 
Using that $m_{33} \sim 0.025$ eV from neutrino oscillation data,  
if $m_h>m_k$ then $\mu\leq \kappa \,m_k$ and $\tilde{I}(r) \sim 1$, so
  eq.~(\ref{eq:numasslimit}) implies that 
\begin{eqnarray}
 &m_h>m_k\gtrsim & \frac{1\, \text{TeV}}{\sqrt{\kappa}} \qquad {\rm NH}, 
\\
 &m_h>m_k \gtrsim & \frac{3 \,  \text{TeV}}{\sqrt{\kappa}} \qquad {\rm IH}.
\end{eqnarray}
On the contrary, if $m_h<m_k$, we find
\begin{eqnarray}
\label{nhlimit}
 &m_k>m_h\gtrsim & \sqrt{\frac{m_k}{m_h \, \kappa \,\tilde{I}(r)}}
\,1\, \text{TeV}\qquad {\rm NH}, 
\\
\label{ihlimit}
 &m_k>m_h \gtrsim &\sqrt{\frac{m_k}{m_h \, \kappa \,\tilde{I}(r)}}
\,3\, \text{TeV} \qquad {\rm IH}.
\end{eqnarray}

From the above results\footnote{Our limits in the IH case differ from those in \cite{AristizabalSierra:2006gb}.
We traced this difference to the fact that in the estimates of \cite{AristizabalSierra:2006gb} 
the perturbativity  bound $|g_{\mu\mu}| < 1$ is imposed, 
but for low  masses, $m_k < 2$ TeV, such bound is always satisfied, and 
the relevant bound is  $|g_{\mu\mu}| \lesssim 0.4 (m_k/ \text{TeV})$, which depends on $m_k$ and changes the scaling with $\epsilon$, leading to a weaker lower bound on the charged scalar masses in our case. We thank Martin Hirsch for discussions about this point.}, we conclude that:

\begin{enumerate}

\item
It is easier to reconcile an  
enhanced Higgs diphoton decay rate with neutrino oscillation data if the former is due to 
the doubly charged scalar loop contribution, since the lower bounds from neutrino masses
are similar, while the BR($H \rightarrow \gamma \gamma$) can be accounted for 
by a heavier $m_k$. Moreover, if the enhancement 
is due to a light $m_h$, then $m_k$ can not be very heavy, because otherwise neutrino masses 
are too small. 

\item
For a NH neutrino mass spectrum, 
  it is  possible to fit simultaneously neutrino oscillation data, lepton flavour violation constraints and an enhanced BR($H \rightarrow \gamma \gamma$) only 
if the trilinear coupling $\mu$ is large, namely 
$\kappa \gtrsim 4 (10)$ for ${\rm min}(m_h,m_k)$ = 500 (300) GeV, 
respectively.

\item
In general, the case of IH neutrino masses is in conflict with an enhanced Higgs diphoton 
rate unless $\kappa \sim {\cal O} (30)$. However if we take into account the strong cancellations
in $\omega_{\mu\mu}$ when $\phi \simeq \delta \simeq \pi$, and allow for a smaller 
$m_{33} \sim 0.003$ eV, it is also possible to fit all data with $ \kappa \sim 4$.

\end{enumerate}

\section{Numerical analysis \label{sec:Numerical-analysis}}

In order to explore exhaustively the highly multi-dimensional parameter space of the ZB model, naive grid scans are completely inappropriate, the method of choice is resorting to Monte Carlo driven Markov Chains (MCMC) that incorporate all the current experimental information described in precedence.
As parameters we will use $\{ s_{ij}^{2},\Delta_{A},\Delta_{S},\phd,\phm,\fmt,m_{h},m_{k},\mu,\gee,\gem,\,\get\}$, and we allow them to vary within the ranges showed in table \ref{tab:values}. 

\noindent Had we tried to use our MCMC to obtain a posteriori probability distribution functions with a canonical Bayesian meaning, the choice of priors would have had a significant role. Nevertheless, since our aim is to explore where in parameter space could the ZB model adequately reproduce experimental data without weighting in the available parameter space volume (that is, the ``metric'' in parameter space given by the priors), we will represent instead profiles of highest likelihood (equivalently profiles of minimal $\chi^2\equiv -2\ln \mathcal L$ with $\mathcal L$ the likelihood) which, on the contrary, can be interpreted on a frequentist basis. This is not a choice that we make because of the merits or demerits of either statistical school: our goal remains to understand if and where the ZB ``works well'', i.e. could fit experimental data. The interpretation of the results/plots will be clear: they show the regions where the model is in agreement with data without regard to their size when the remaining information (parameters and observables) is marginalized over\footnote{Typically both approaches should converge to similar results when (experimental) information abounds; in a study such as this one, if they differ, rather than sticking to one or the other, from the physical point of view we would only conclude that the current experimental data is not yet sufficient to pin down or exclude the model.}.
In this case, exploring the parameter space in a uniform, logarithmic or other manner, in some given parameter will not affect our results (only the computational efficiency required to reach them will be, of course, affected). 

For the modelling of experimental data we typically resort to individual Gaussian likelihoods for measured quantities.
Bounds are implemented through smooth likelihood functions that include, piecewise, a constant and a Gaussian-like behaviour. For the sake of clarity: if the experimental bound for a given observable $\mathcal O$ is $B^{\mathcal O}_{[\text{90\%CL}]}$ at 90\% CL (1.64$\sigma$ in one dimension), the $\chi^2$ contribution associated to the model prediction $\mathcal O_{\rm th}$ for this observable is
\[
\chi^2(\mathcal O_{\rm th})=\left\{\begin{array}{l}0,\  \mathcal O_{\rm th}<B^{\mathcal O}_{[\text{90\%CL}]}/1.64,\\ \left(\frac{1.64\mathcal O_{\rm th}}{B^{\mathcal O}_{[\text{90\%CL}]}}-1\right)^2\left(\frac{1.64}{0.64}\right)^2,\ \mathcal O_{\rm th}\geq B^{\mathcal O}_{[\text{90\%CL}]}/1.64.\end{array}\right.
\]
In this way we avoid imposing sharp stepwise bounds or half-Gaussian with best value at zero that may penalize deviating from null predictions when this might not be supported by experimental evidence (in particular when the number of bounds included in the analysis is significant). 

Simulations are done for both normal and inverted hierarchy. In each point of the 
parameter space we compute the full
$\chi^2$, including all measurements and bounds. 
In the  plots we show the regions with the total $\Delta \chi^2 \leq 6$, which corresponds to 95\% confidence levels with two variables.

\begin{table}[ht]
\begin{centering}\begin{tabular}{||c|c|c||}
\hline 
Parameter &
Allowed range \\
\hline
\hline 
$\Delta_{S}$ &
\, $(7.50\pm0.19)\times10^{-5}\,\mathrm{eV^{2}}$\, \\
\hline 
$\Delta_{A}$ &
$(2.45 \pm 0.07)\times10^{-3} \mathrm{eV^{2}}$\\ 
\hline 
$\sin^{2}\theta_{12}$ &
$0.30\pm0.13$\\
\hline 
$\sin^{2}\theta_{23}$ &
$(0.42\pm0.04) \cup (0.60\pm 0.04)$\\ 
\hline 
$\sin^{2}\theta_{13}$ &
$0.023\pm0.002$\\ 
\hline
$\delta, \phi$ &
$[0, 2\pi]$\\ 
\hline 
\, $\arg(g_{ee}),\arg(g_{e\mu}),\arg(g_{e\tau})$ \, &
$[0, 2\pi]$ \\ 
\hline 
$f_{\mu\tau},|g_{ee}|,|g_{e\mu}|,|g_{e\tau}|$ &
$[10^{-7}, 5]$\\ 
\hline 
$m_{h}$ &
$[100,  2 \times10^{3}]\,\mathrm{GeV}$\\
\hline 
$m_k$ &
$[200,  2 \times10^{3}]\,\mathrm{GeV}$\\
\hline 
$\mu$ &
$[1, 2\kappa\times10^{3}]\,\mathrm{GeV}$\\ 
\hline 
\end{tabular}\par\end{centering}
\caption{Allowed ranges  for the parameter scan (Neutrino oscillation parameters are obtained from \cite{GonzalezGarcia:2012sz,Tortola:2012te,Fogli:2012ua}).}
\label{tab:values} 
\end{table}

To compare our results with the analysis presented a few years ago by some of us \cite{Nebot:2007bc} some remarks 
are in order: first, here we have updated the experimental input on LFV and neutrino oscillation parameters, as well as LHC direct searches.
The new limits, in particular on $\mu \rightarrow e\gamma$, 
 tend to reduce the allowed regions but not dramatically. 
Especially important is the determination of $\sin\theta_{13}$: as shown in 
\cite{Nebot:2007bc}, already before its measurement the ZB model predicted a large 
mixing angle $\theta_{13}$ in the case of IH spectrum, close to the previous experimental upper limit, while for NH any value of $\theta_{13}$  below the bound was allowed.
In fact, a very small value of $\theta_{13}$ 
would have ruled out the IH possibility within the ZB model.
Second, although the scanning of parameters is performed like in \cite{Nebot:2007bc}, we have chosen here to present results in terms of profiles of highest likelihood,
which are insensitive to the volume of the parameter space and the priors used to scan it. This allows us to explore
regions where parameters are fine tuned (after all, Yukawa couplings always require a certain degree of fine tuning). 
This is important since, as we have seen, the model is highly constrained at present and less conservative assumptions
could exclude it before time, at least in the region of low masses. Moreover, we focus only on the region of
masses with phenomenological interest ($m_{h,k} < 2$~TeV) precisely to explore better the region of low masses.

\begin{figure}
	\centering
	\includegraphics[width=0.45\textwidth]{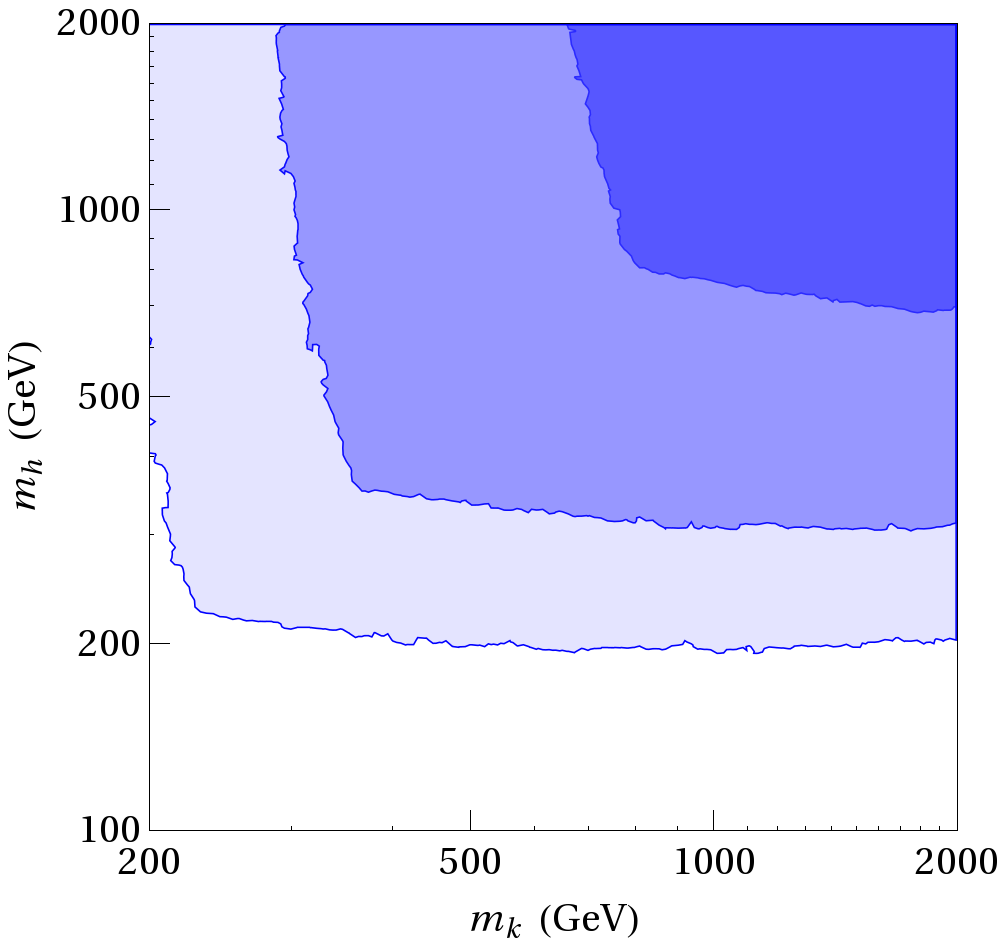}~~
	\includegraphics[width=0.45\textwidth]{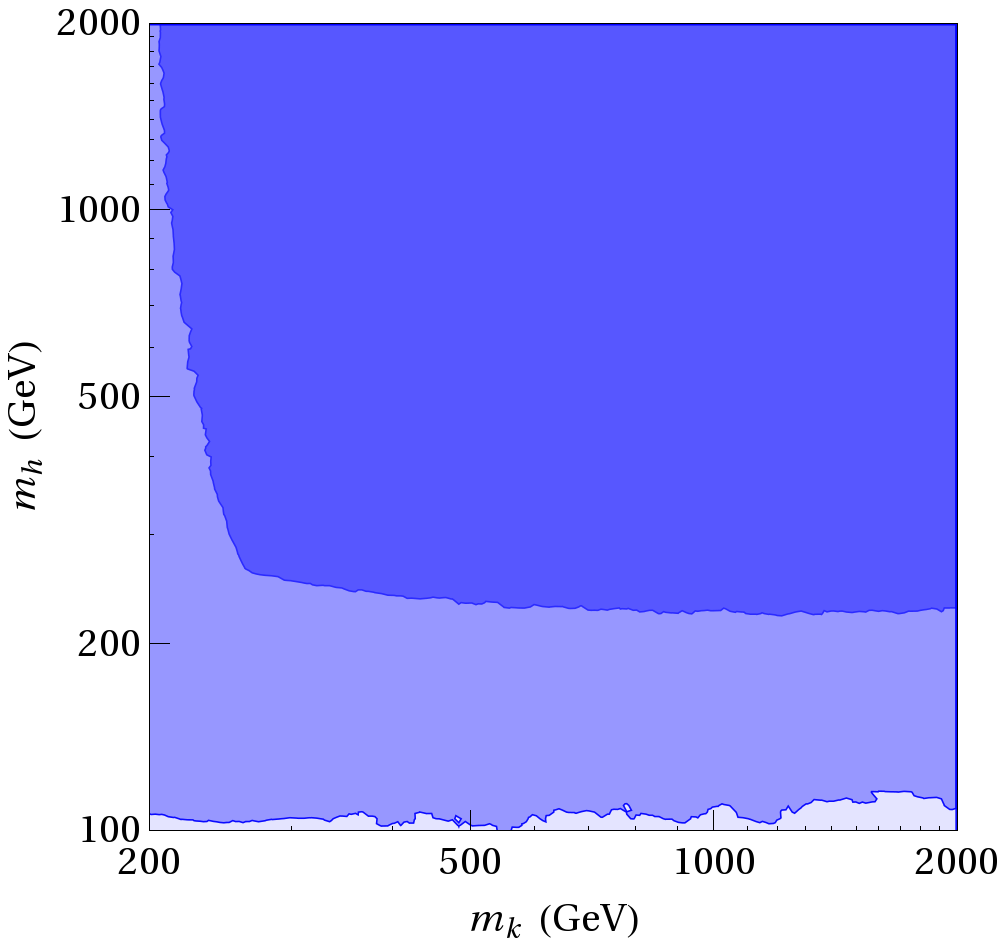}
	\caption{$m_h$ vs $m_k$ for NH (left) and IH (right) for different values of the perturbative parameter $\kappa=1,5,4\pi$ (dark to light colours).} \label{fig:masas}
\end{figure}

\begin{figure}
	\centering
	\includegraphics[width=0.45\textwidth]{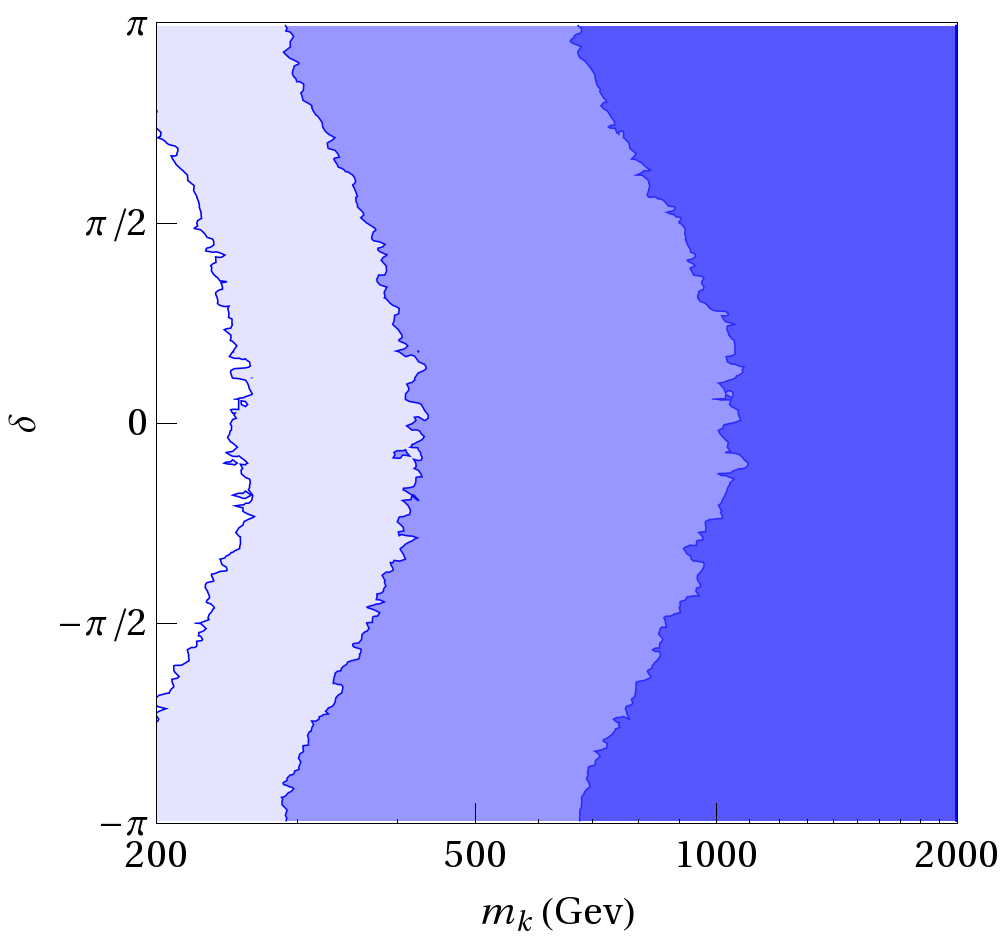}
	\includegraphics[width=0.45\textwidth]{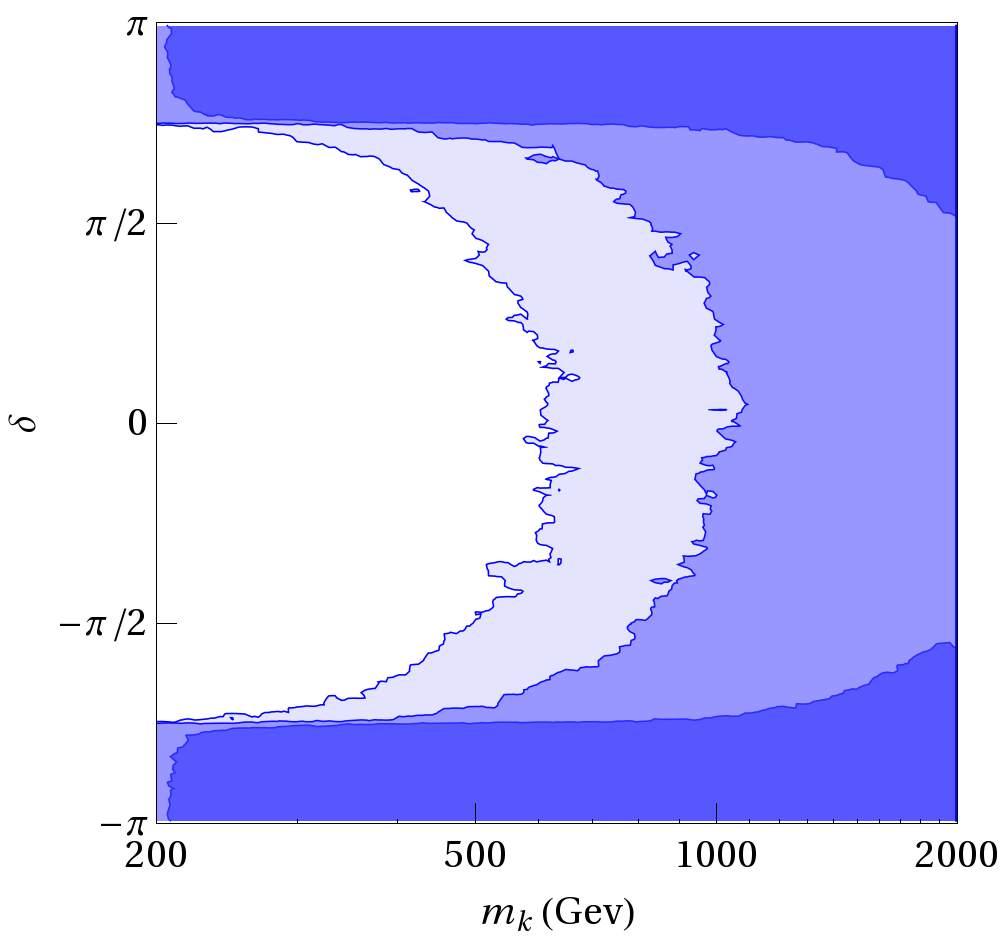}
	\caption{$\delta$ vs $m_k$ in NH (left) and IH (right).} \label{fig:deltam}

\end{figure}

In fig.~\ref{fig:masas} we depict the points allowed by neutrino oscillation data and all low energy constraints in the plane $(m_h, m_k)$ for the two mass orderings
(NH and IH) and different values of the fine-tuning parameter in eq.~\eqref{mu} ($\kappa=1$ darker, $\kappa=5$ dark, $\kappa=4\pi$ light). 
The results of the numerical analysis imply that in general the indirect lower bounds on $m_h$ and $m_k$ from neutrino oscillation data and low energy constraints are stronger than the current limits from direct searches, except when cancellations occur for 
$\delta,\phi \sim \pi$, especially in the IH case, and/or when naturality assumptions on $\mu$ are relaxed, allowing for $\kappa=4 \pi$.
In table~\ref{tab:scalar-masses} we summarize the 
lower bounds on the scalar masses obtained for the three values of the naturality parameter 
$\kappa$, and two illustrative values of the Dirac phase, $\delta=0, \pi$. For $\delta \sim -\pi/2$, as might be suggested by a recent analysis \cite{GonzalezGarcia:2012sz}, the bounds are slightly weaker than in the $\delta=0$ case (see fig.~\ref{fig:deltam}).

\begin{table}
\begin{tabular}{|c|c|c|c|c|c|c|}
\hline 
 & \multicolumn{1}{c}{} & \multicolumn{1}{c}{NH} &  & \multicolumn{1}{c}{} & \multicolumn{1}{c}{IH} & \tabularnewline
\hline 
\hline 
$\kappa$ & $1$ & $5$ & $4\pi$ & $1$ & $5$ & $4\pi$\tabularnewline
\hline 
$m_{h}\,(\mathrm{GeV)}$ & $700\,(1000)$ & $300\,(400)$ & $200\,(250)$ & $220\,(>2000)$ & $100\,(1000)$ & $100\,(650)$\tabularnewline
\hline 
$m_{k}\,(\mathrm{GeV)}$ & $700\,(1100)$ & $300\,(450)$ & $200\,(250)$ & $200\,(>2000)$ & $200\,(1000)$ & $200\,(550)$\tabularnewline
\hline 
\end{tabular}
\caption{Lower bounds for the scalar masses for NH and IH and the naturality constraints parametrized by the three values of $\kappa$. We present results for $\delta=\pi$ ($\delta=0$) (see figs.~\ref{fig:masas} and \ref{fig:deltam}).\label{tab:scalar-masses}}
\end{table}

The correlation between the CP phase $\delta$ of the
neutrino mixing matrix and the scalar masses is illustrated in 
fig.~\ref{fig:deltam}, where we plot $\delta$ versus the 
doubly charged scalar mass, $m_k$.\footnote{The correlation of $\delta$ with 
$m_h$  is entirely analogous.}  Such correlation is especially relevant 
  in the IH case, where  scalar masses lower than 
$\sim$ 1 TeV are only allowed if $\delta \sim \pi$.  
A similar correlation with the phase $\phi$ was already found in \cite{Nebot:2007bc}
for IH spectrum,  so we do not show it here.

\begin{figure}
\begin{centering}
\includegraphics[width=0.42\columnwidth]{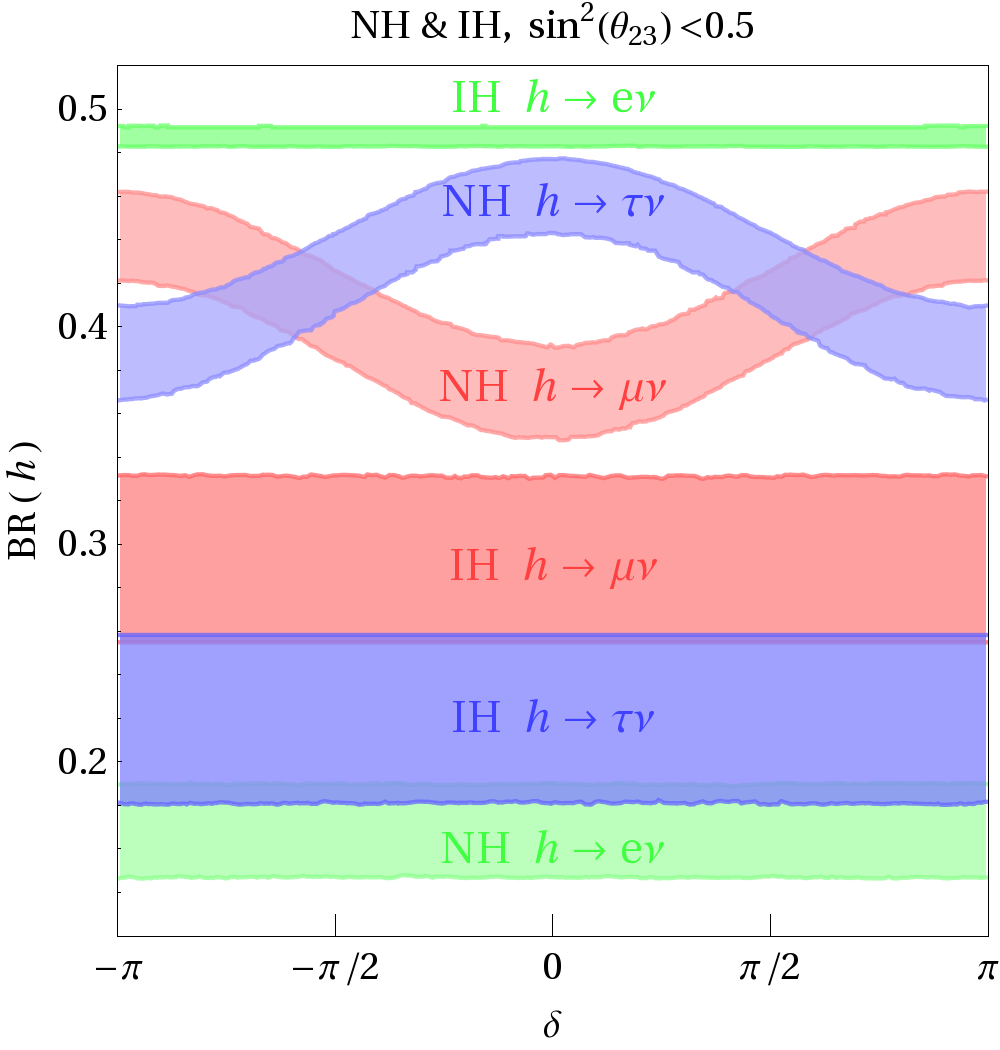}~~
\includegraphics[width=0.42\columnwidth]{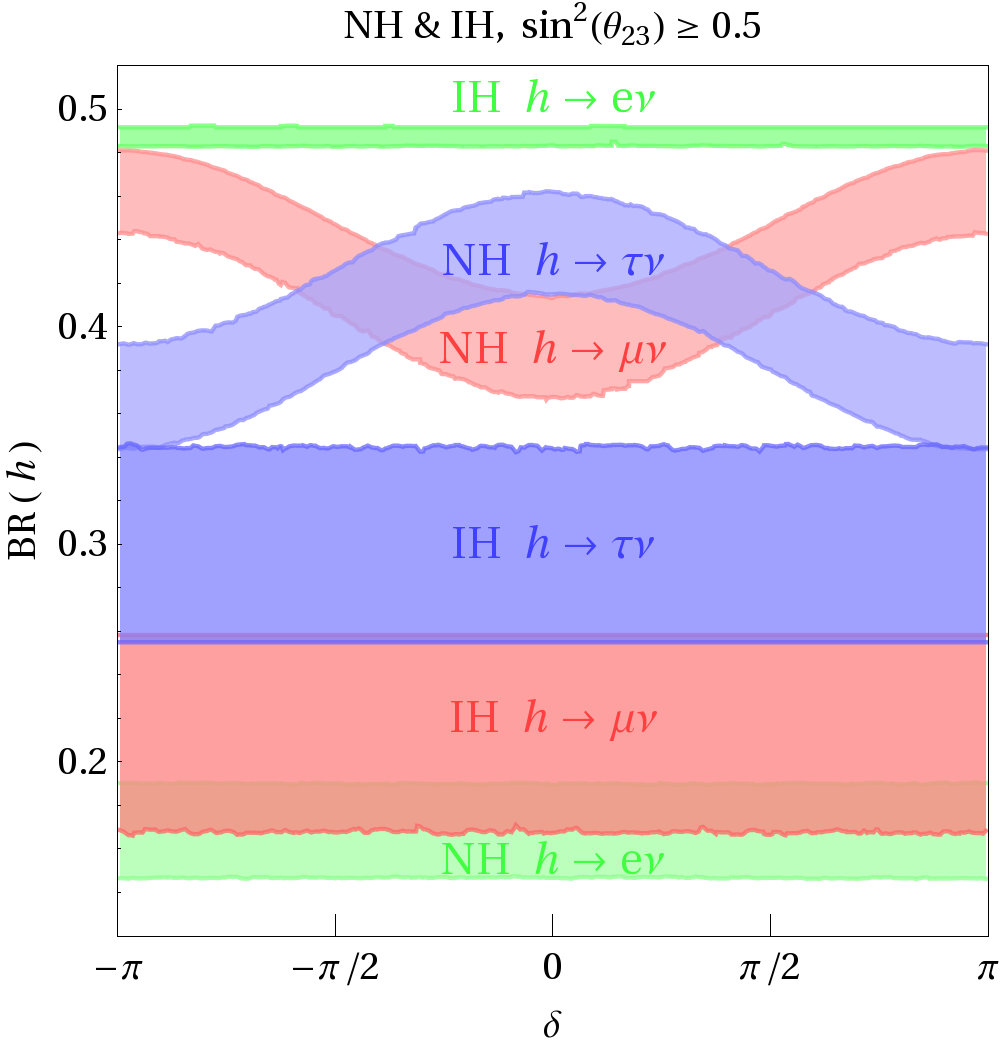}
\par\end{centering}
\caption{Branching ratios of the charged singlet $h$ to $e\nu, \mu\nu,\tau\nu$ splitting the two currently allowed octants of $\theta_{23}$, $\theta_{23} < 45^\circ$ ($\theta_{23} > 45^\circ$) left (right). One can see the dependence on $\delta$ for the NH spectrum in the $\mu\nu$ and $\tau\nu$ channels. The most significant change between octants is the interchange of the $\mu\nu$ and $\tau\nu$ for the IH case. The bands are $95$\% C.L. regions.}
\label{fig:h}
\end{figure}

Regarding the singly charged scalar $h^{\pm}$, the width of its 
 decay modes
($e\nu, \mu\nu,\tau\nu$) is fixed by
the $f_{ia}$ couplings to leptons (see for instance \cite{AristizabalSierra:2006gb, Nebot:2007bc} for the relevant formulae).
Therefore, after the measurement of $\theta_{13}$, present neutrino oscillation data determine completely the BRs of $h$  from eqs. \eqref{fnh} and \eqref{fih},  
up to a  residual dependence on the CP phase $\delta$ in the case of NH spectrum. 
In this case, a very precise measurement of the branching ratios in the $\mu\nu$ or $\tau \nu$ channels (probably in a next generation collider) will predict the CP phase $\delta$, and viceversa.
We show the ranges attainable by the different BRs in fig.~\ref{fig:h}, as a function of $\delta$, 
splitting the two currently allowed octants of $\theta_{23}$.
The most significant change between octants is the interchange of the $\mu\nu$ and $\tau\nu$ for the IH case.
Clearly, the best option to discriminate between hierarchies is the $e\nu$ channel.

An important point of the ZB model is that the doubly charged scalar can decay to two singly charged scalars, which are difficult to detect at the LHC. However, 
in fig.~\ref{fig:masas} we see that for a NH neutrino mass spectrum
 $m_h>200$ GeV, and the channel $k\rightarrow hh$ is closed for $m_k<400$ GeV. Therefore, present bounds on $m_k$ from dilepton searches at LHC discussed in \ref{sec:LHC} apply. 
 For the IH case, the $k \rightarrow h h $ channel is always open and can be dominant, unless $\kappa=1$, for which we obtain that it is closed in the region $m_k<440$ GeV. 
Thus in general current direct  bounds from LHC are weaker.

\begin{figure}
	\centering
	\includegraphics[width=0.44\textwidth]{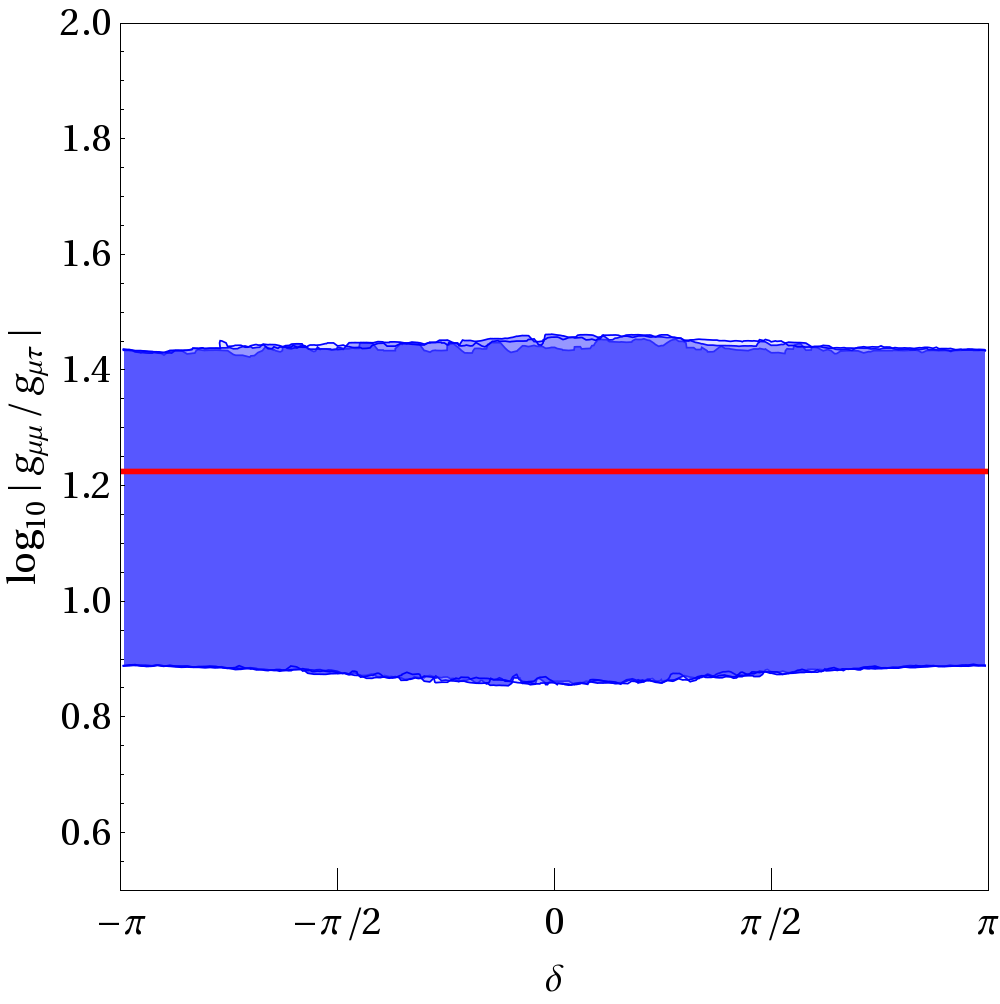}~~
	\includegraphics[width=0.44\textwidth]{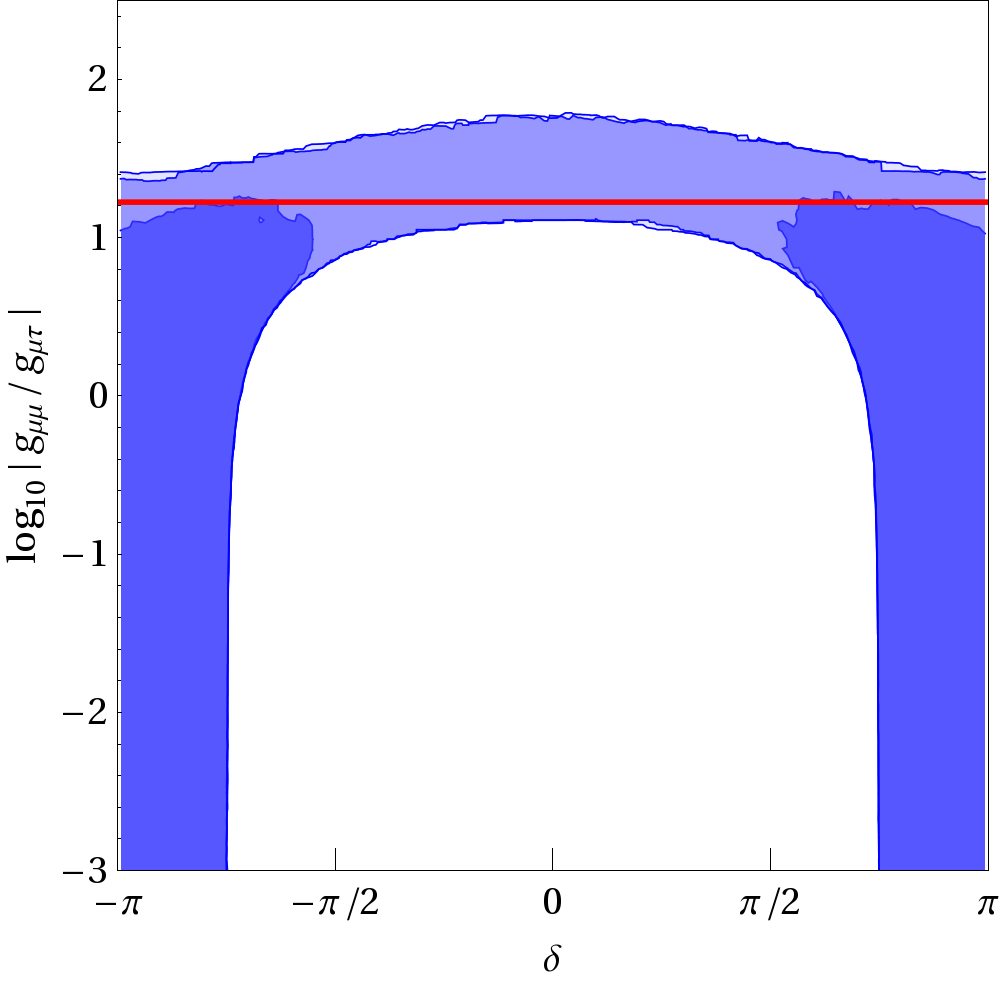}
	\includegraphics[width=0.46\textwidth]{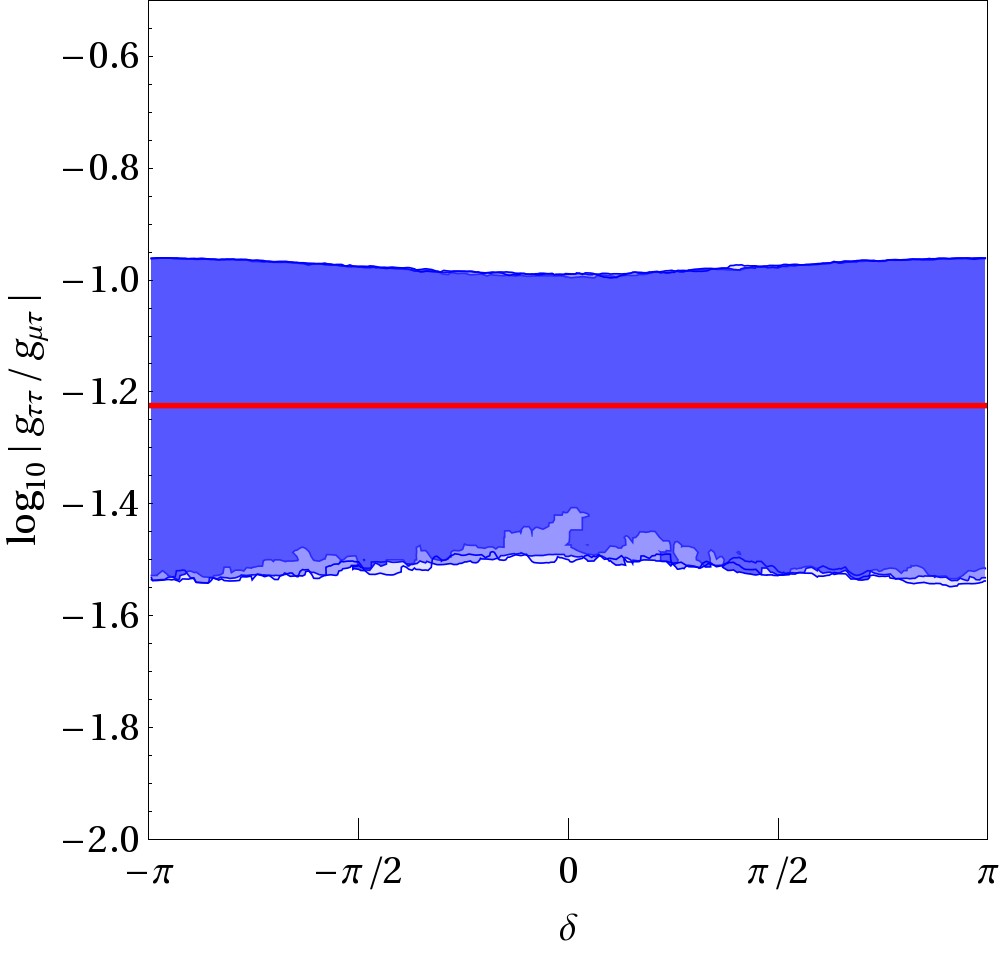}~~
	\includegraphics[width=0.46\textwidth]{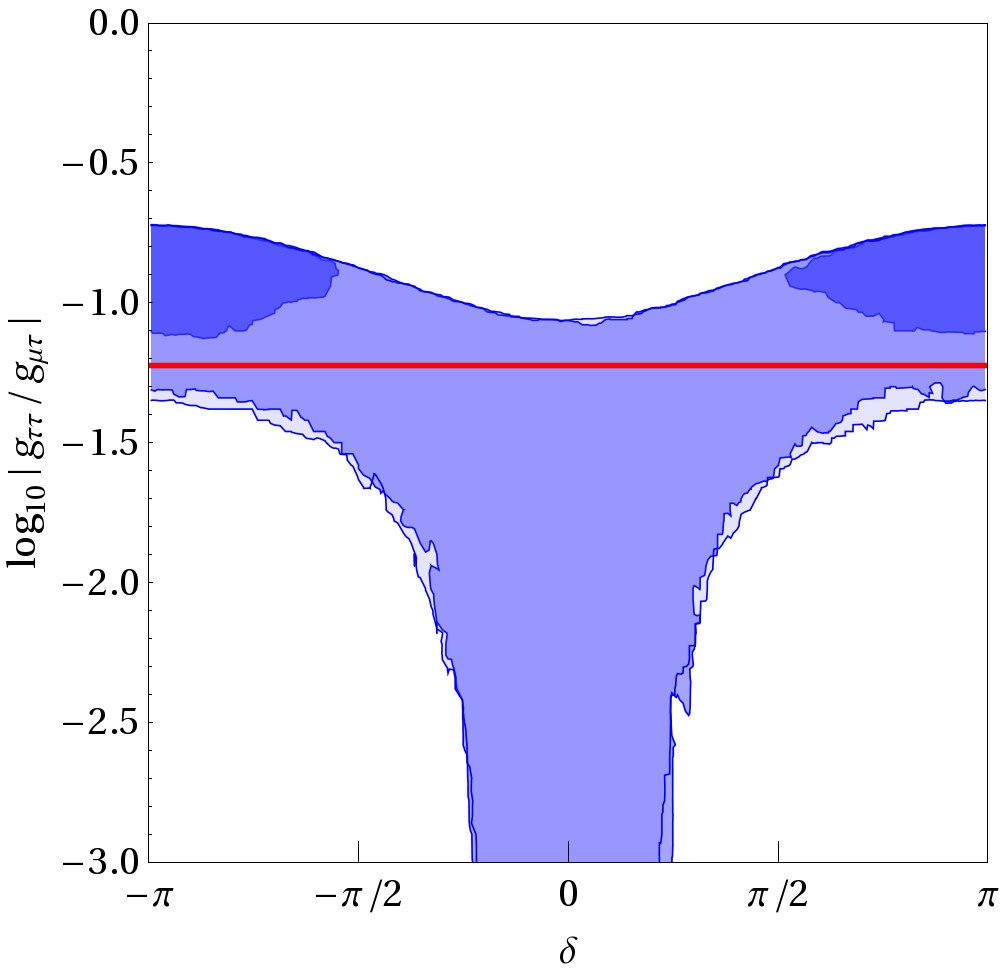}
	\caption{$\log|g_{\mu\mu}/g_{\mu\tau}|$ and $\log|g_{\tau\tau}/g_{\mu\tau}|$ vs $\delta$ for NH (left) and IH (right). The horizontal red lines represent the naive approximation in eq.~\eqref{couplinghierarchy}. \label{fig:gratios}}
\end{figure}

Let us now turn to the $g_{ab}$ couplings. We find  always $g_{\tau\tau} \ll g_{\mu\tau}$, both for the NH and IH cases, in agreement with the analytic estimates in eq.~(\ref{couplinghierarchy}); however the expected ratio 
$g_{\mu\mu} / g_{\mu\tau} \sim m_{\tau}/m_{\mu}$
is only fulfilled for the NH spectrum, since in the IH case large cancellations when the phases
of the PMNS matrix $U$ are $\delta \sim \phi \sim \pi$ lead to smaller 
$g_{\mu\mu}  \ll g_{\mu\tau}$. 
This can be seen in fig.~\ref{fig:gratios}, where we show the ratios  
$g_{\tau\tau}/g_{\mu\tau}$  and $g_{\mu\mu}/g_{\mu\tau}$ 
obtained in the numerical simulation as a function of $\delta$,
together with the expectation based on the analytic approximations, which is just a constant
fixed by the charged lepton masses (red horizontal line)\footnote{In the NH case there can also be cancellations with the $g_{e\tau}$ terms, which have been neglected in eq.~\eqref{eq:mijNangulos}, that would allow much smaller values of $g_{\tau \tau}$ and $g_{\mu \tau}$, but those only occur for $\kappa=4\pi$ and in a tiny region of the parameter space.}.

\begin{figure}
	\centering
	\includegraphics[width=0.47\textwidth]{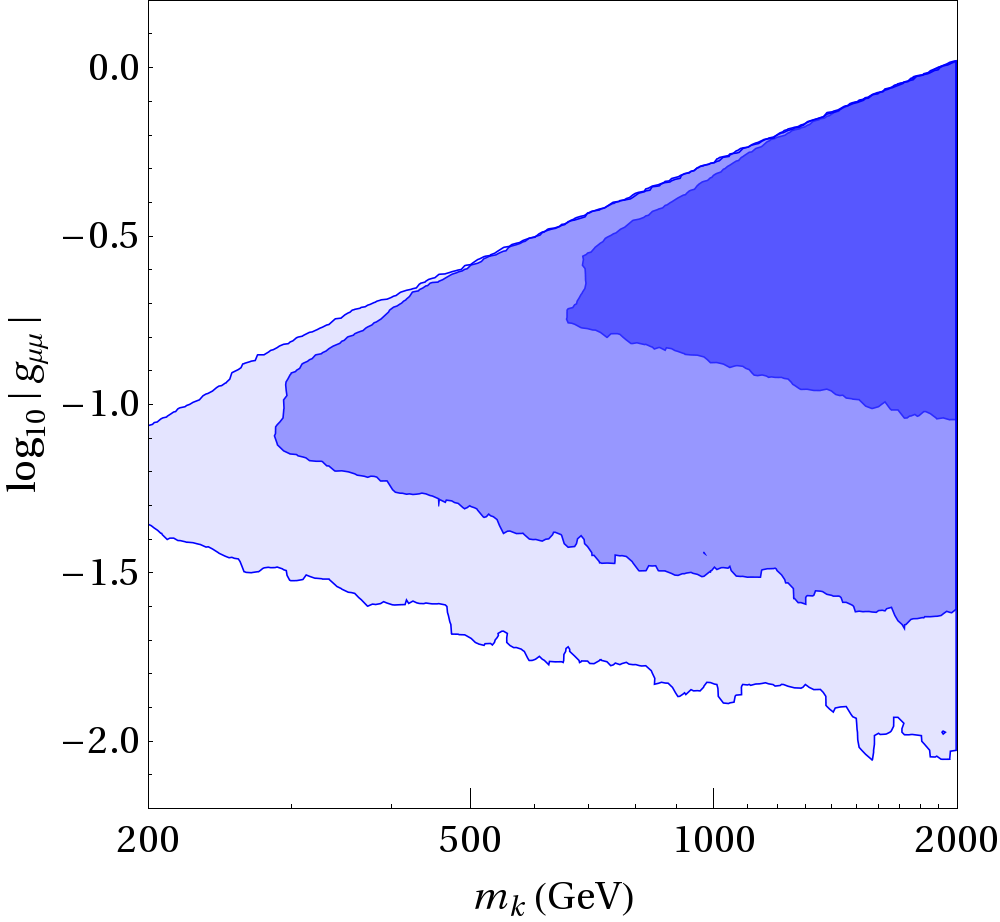}~~
	\includegraphics[width=0.47\textwidth]{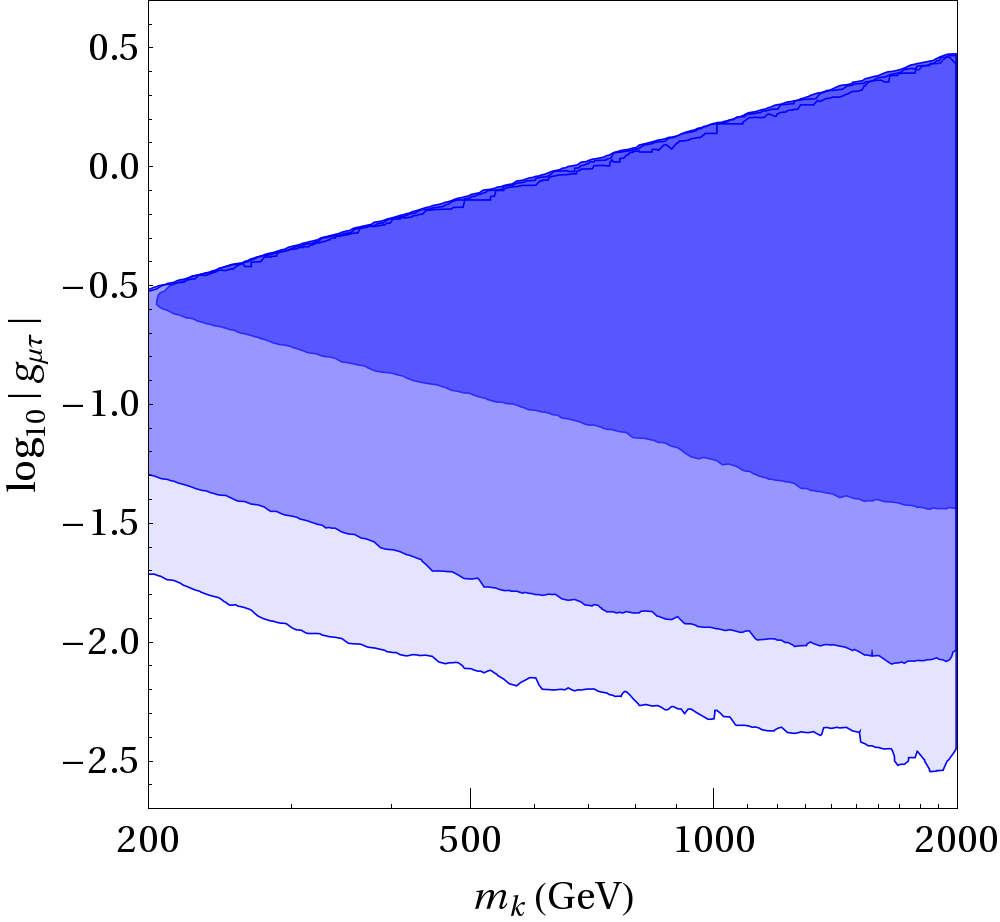}
	\caption{$\log|g_{\mu\mu}|$ vs $m_k$ for NH (left) and $\log|g_{\mu\tau}|$ vs $m_k$ for IH (right).} \label{fig:gmmvsmk}
\end{figure}

To set the absolute scale of the couplings we present in fig.~\ref{fig:gmmvsmk} the value of the largest couplings against $m_k$, namely 
$g_{\mu\mu}$ in the NH case, and $g_{\mu\tau}$ in the IH case. 
We see that  in both cases the couplings are always in the range from $10^{-2}$ to $1$ and therefore they tend to dominate the decays of the $k^{++}$. 

\begin{figure}
	\centering
	\includegraphics[width=0.45\textwidth]{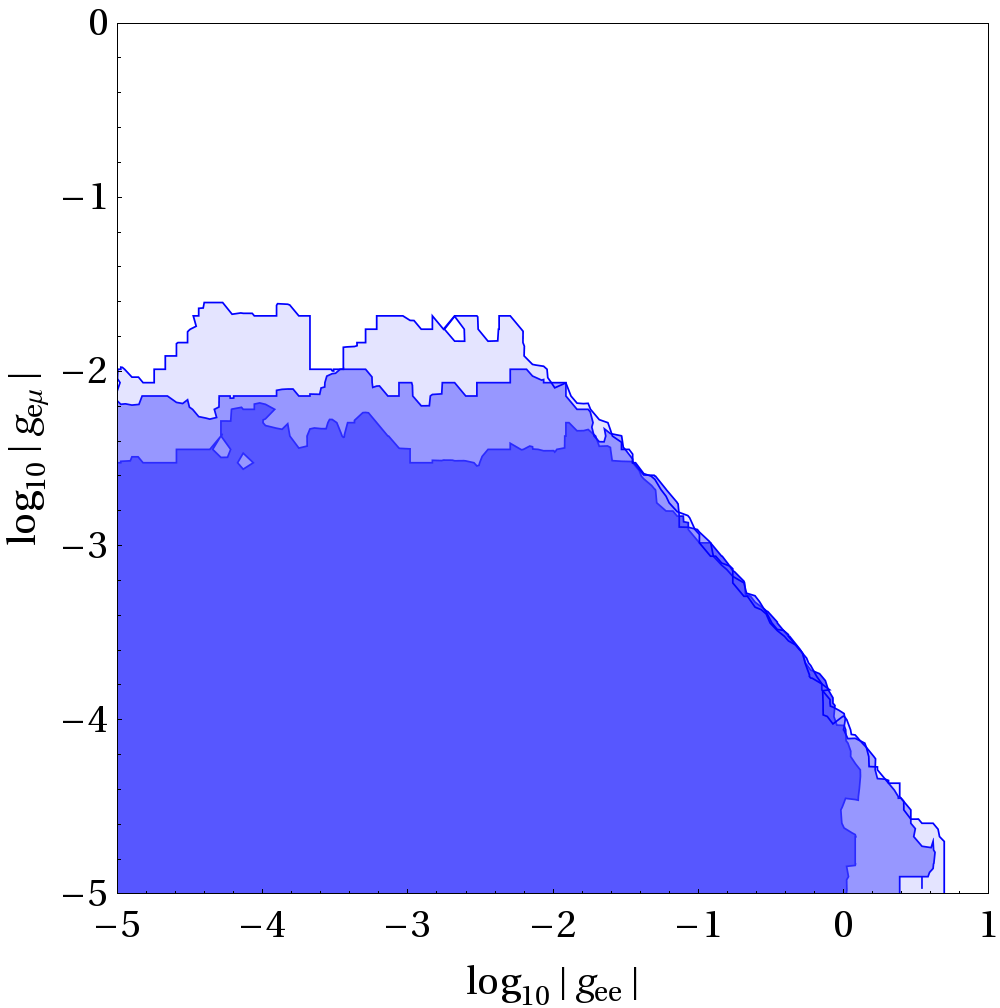}~~
	\includegraphics[width=0.45\textwidth]{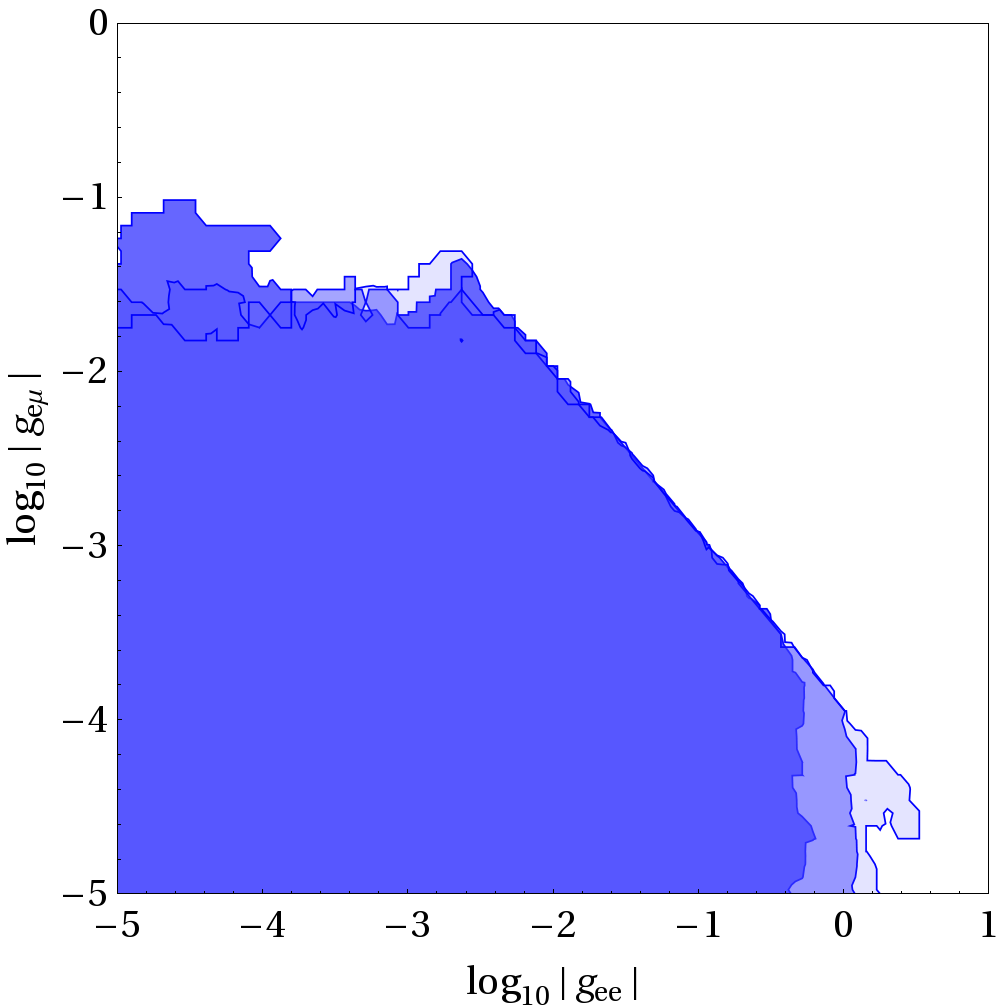}
	\caption{$\log|g_{e\mu}|$ vs $\log|g_{ee}|$ for NH (left) and IH (right).} \label{fig:gem-gee}
\end{figure}

Regarding 
the couplings $g_{ea}$, which  are not determined by the neutrino mass matrix, 
bounds from LFV charged lepton decays strongly constrain $g_{e\tau}$  
and $g_{e\mu}$  
to be less than ${\cal O}$(0.01), while $g_{ee}$ can be larger, ${\cal O}$(1).
The constraint on 
$|g_{ee} g_{e\mu}|$  from $\mu\rightarrow 3e$  implies that 
$|g_{ee} g_{e\mu}|<2.3\times10^{-5}\,(m_{k}/ \mathrm{TeV})^{2}$ and 
it is illustrated in fig.~\ref{fig:gem-gee}. 

Since the widths of the $k^{\pm\pm}$ leptonic decay modes are directly related to these couplings, 
from the above results we can readily infer the corresponding BRs. 
We find that the probability 
of $k \rightarrow e \mu, e\tau, \tau\tau$ is always negligible (even in the IH case, $g_{e\mu}$ can be at most $0.1$ and only when $\delta \sim \pi$).
 For $m_k \lesssim 400$ GeV, and NH neutrino spectrum, 
BR($k \rightarrow ee$) + BR($k \rightarrow \mu\mu) \sim 1$, 
since the $k \rightarrow h h$ decay channel is closed; 
therefore $k^{\pm\pm}$ can not
evade current LHC bounds on doubly-charged scalar searches and the  limit 
$m_k > 310$ GeV applies ($400$ GeV if no signal is found at 8 TeV with 20 $\text{fb}^{-1}$
\cite{delAguila:2013mia}). 
In the same $m_k$ range, for IH neutrino spectrum the 
BR($k \rightarrow \mu\tau)$ can also be significant and 
the  channel $k \rightarrow h h$ is open (unless $\kappa =1$, for which 
it is only open for $m_k > 440$ GeV),  thus the present bound is weaker.

When  the upcoming LHC 13-14 TeV  data is available, it is important to take into account that the decay channel $k \rightarrow h h$ is open for $m_k \gtrsim 400$ GeV, and
can be dominant, so in this mass range limits on doubly-charged scalars from dilepton searches will not apply to  the ZB model.
On the contrary, 
 if a doubly charged scalar were detected at LHC in any mass range, neutrino oscillation data and low energy constraints are powerful enough to falsify the ZB model to a large extent. For instance,  we know that 
BR($k \rightarrow e\mu,e\tau,\tau\tau$) are negligible for 
any neutrino mass spectrum, while a sizeable BR($k \rightarrow \mu\tau$) is only compatible with 
an IH spectrum. 

\section{Conclusions \label{sec:Conclusions}}

We have analyzed the ZB model in the light of recent data: the measured neutrino mixing angle 
$\theta_{13}$, limits from the rare decay $\mu\rightarrow e \gamma$ and LHC results. 
Although the model contains many free parameters, neutrino oscillation data and low energy constraints are powerful enough to rule out sizeable  regions of the parameter space. 
A large source of uncertainty comes from the mass scale of the new physics, which is 
unknown. Since we are interested on possible signatures at the LHC, we present results for the masses
of the extra scalar fields below 2 TeV. 
Previous analyses \cite{AristizabalSierra:2006gb,Nebot:2007bc} have shown that larger mass scales are always 
allowed,  given the absence of significant deviations from the SM besides neutrino masses.

Even within this reduced scenario, there is still a free mass parameter, the trilinear 
coupling between the charged scalars, $\mu$, which remains mainly unconstrained.   
Naturality arguments together with perturbativity and vacuum stability bounds, indicate 
that $\mu$ can not be much larger than the physical scalar masses, $m_k,m_h$, but it is not possible to determine a precise theoretical limit.  
Because the neutrino masses depend linearly on the parameter $\mu$, the ability of the model to accommodate all present data is  quite sensitive to the upper limit allowed for it, so 
we have considered three limiting values, $\mu< \kappa \,  {\rm min}(m_k,m_h)$, 
with $\kappa = 1, 5, 4\pi$. 
 Within the above ranges for the mass parameters of the ZB model, we have 
performed an exhaustive numerical analysis using Monte Carlo Markov Chains (MCMC), 
incorporating all the current experimental information available, both for NH and IH neutrino 
masses.  The results of the analysis are presented in sec.~\ref{sec:Numerical-analysis} and summarized in figs. \ref{fig:masas} -- \ref{fig:gem-gee}.

We have addressed  the possibility that the slight excess in the Higgs diphoton decay 
observed by the ATLAS collaboration is due to  virtual loops of the extra charged scalars 
of the ZB model, $h^\pm$ and $k^{\pm\pm}$. Note that in the Zee-Babu model, as the new particles are singlets, there is a negative correlation between $H \rightarrow \gamma \gamma$ and $H \rightarrow \gamma Z$. Although a similar study has been performed in
\cite{Chao:2012xt}, it was limited to the scalar sector parameters of the model, and neutrino data, which we 
find crucial to determine the allowed charged scalar masses, was not included in the analysis.  
In agreement with \cite{Chao:2012xt}, we find that in order to accommodate an enhanced 
$H \rightarrow \gamma \gamma$ decay rate, large and negative 
$\lambda_{hH}, \lambda_{kH}$ couplings are needed, together with light scalar masses
$m_h < 200$ GeV, $m_k < 300$ GeV. Such couplings are in conflict with the stability of the 
potential, unless the self-couplings $\lambda_{h,k}$ are pushed close to the naive perturbative 
limit, $\sim 4 \pi$. As a consequence, even if vacuum stability and perturbativity constraints are 
satisfied at the electroweak scale, RGE running leads to non-perturbative couplings 
at scales not far from the electroweak scale, as shown in fig.~\ref{fig:perturbativity}. 

When neutrino data and low energy constraints are taken into account,
we still find regions of the parameter space in which such enhancement is compatible with 
all current experimental data; 
in particular, it seems easier if the enhancement is due to  the doubly-charged scalar 
loop contribution. As can be seen in fig.~\ref{fig:masas}, 
in the NH case, the trilinear coupling $\mu$ should be near its upper limit, 
while in the IH case lower masses can be achieved in the region $\delta \sim \phi \sim \pi$
due to cancellations. 

Regarding LHC bounds on the doubly-charged scalar mass, they are largely dependent on the 
BRs of the $k^{\pm\pm}$ decay modes, namely same sign leptons $\ell^\pm_a \ell^\pm_b$ and 
$h^\pm h^\pm$. The leptonic decay widths are controlled by the $g_{ab}$ couplings to the right-handed leptons, which are in principle unknown. 
By imposing that the measured neutrino mass matrix is reproduced,  
within the approximation $m_e = 0$ one obtains 
analytically that  
$g_{\tau\tau}:g_{\mu\tau}:g_{\mu\mu}\sim m_{\mu}^{2}/m_{\tau}^{2}:m_{\mu}/m_{\tau}:1$, 
while there is no information on the  $g_{ea}$ couplings. 
Our numerical analysis confirms the above ratio of couplings in the case of NH, but for 
the IH spectrum there can be large cancellations if the PMNS matrix phases $\delta,\phi$
are close to $\pi$, leading to $g_{\tau\tau} \ll g_{\mu\tau} \sim g_{\mu\mu}$. 
In both cases, $g_{e\mu}, g_{e\tau} \lesssim 0.1$. 

Moreover,  in NH, if  $m_k < 400$ GeV for $\kappa = 4 \pi$
($m_k < 600$ GeV if $\kappa =5$),  $m_h < m_k/2$ is ruled out, 
therefore the decay channel  $k \rightarrow h h $ is kinematically closed 
and the LHC bounds from doubly-charged scalar searches can not be evaded.
 In IH, however, for $\delta \sim \phi \sim \pi$ the $k \rightarrow h h $ channel is open unless 
 $\kappa=1$,
while if $\delta$ is very different from $\pi$, indirect bounds on $m_k$ set a much stronger constraint than direct LHC searches.

As a  consequence, if the light neutrino spectrum is NH, $k$ decays mainly to $ee,\mu\mu$,
and the current bound from LHC is $m_k > 310$ GeV, while if the spectrum is IH, 
$k$ may also decay to $\mu\tau$ and $hh$, so the present bound is weaker,  about 200 GeV. 
Were a doubly-charged boson discovered at LHC, the measurement of its leptonic BRs could 
rule out the ZB model, or predict a definite neutrino mass spectrum.
Conversely, if a CP phase $\delta$ is measured in future neutrino oscillation experiments to be quite different from $\pi$ 
together with an IH spectrum, the mass of the charged scalars of the ZB model will be pushed up well outside the 
LHC reach. 

{\bf Note:} During the final stages of this work we became aware of \cite{Schmidt:2014zoa}, where an analysis of the Zee-Babu model was performed. 
Our bounds on the scalar masses are comparable to theirs taking into account the slightly different procedures, in particular that   they fix the neutrino oscillation parameters to their best fit values and we allow them to vary in their two sigma range.  
While in our work we focus on prospects for the LHC, in \cite{Schmidt:2014zoa}  the possibility of detecting the doubly charged singlet in a future linear collider is studied.

\begin{acknowledgments}
We are thankful to the authors of \cite{Schmidt:2014zoa} for sharing with us their work and for useful discussions. 
We also thank Marcela Carena, Ian Low and Carlos Wagner for discussions. 
This work has been partially supported by the  European Union FP7  ITN INVISIBLES (Marie Curie Actions, PITN- GA-2011- 289442), 
by the Spanish MINECO under grants  FPA2011-23897, FPA2011-29678, 
Consolider-Ingenio PAU (CSD2007-00060)
and CPAN (CSD2007- 00042) and by Generalitat Valenciana grants PROMETEO/2009/116
and PROMETEO/2009/128.  M.N. is supported by a postdoctoral fellowship of project CERN/FP/123580/2011 at CFTP (PEst-OE/FIS/UI0777/2013), projects granted by \emph{Funda\c{c}\~ao para a Ci\`encia e a Tecnologia} (Portugal), and partially funded by POCTI (FEDER).
J.H.-G. is supported by the MINECO under the FPU program. He would also like to acknowledge NORDITA for their hospitality during the revision of the final version of this work, carried out during the ``News in Neutrino Physics" program.

\end{acknowledgments}

\appendix

\section{RGEs in the ZB model} \label{RGE}
\begin{eqnarray}
\label{eq:RGES}
16 \pi^2 \beta_H &=& \frac 3 8 \left[(g^2+g'^2)^2 + 2 g^4 \right] - (3 g'^2 + 9 g^2) \lambda_H
+24 \lambda_{H}^2 + \lambda_{hH}^2 + \lambda_{kH}^2 - 6 y_t^4 + 12 \lambda_H y_t^2
\nonumber \\
16 \pi^2 \beta_h &=& 6g'^4  - 12 g'^2 \lambda_h
+ 20 \lambda_{h}^2 + 2 \lambda_{hH}^2 + \lambda_{hk}^2
\nonumber \\
16 \pi^2 \beta_k &=& 96 g'^4 - 48 g'^2 \lambda_k
+ 20 \lambda_{k}^2 + 2 \lambda_{kH}^2 + \lambda_{hk}^2
\nonumber \\
16 \pi^2 \beta_{hH} &=& 3g'^4 - (\frac {15}{2} g'^2 + \frac 9 2 g^2) \lambda_{hH}
+12 \lambda_H \lambda_{hH} + 8 \lambda_h  \lambda_{hH}  + 
2 \lambda_{kH} \lambda_{hk} + 4 \lambda_{hH}^2  + 6 \lambda_{hH} y_t^2
\nonumber \\
16 \pi^2 \beta_{kH} &=& 12 g'^4 - (\frac {51}{2} g'^2 + \frac 9 2 g^2) \lambda_{kH}
+12 \lambda_H \lambda_{kH} + 8 \lambda_k  \lambda_{kH}  + 
2 \lambda_{hH} \lambda_{hk} + 4 \lambda_{kH}^2  + 6 \lambda_{kH}  y_t^2
\nonumber \\
16 \pi^2 \beta_{hk} &=& 48 g'^4 - 30 g'^2  \lambda_{hk}
+4 \lambda_{kH} \lambda_{hH} + 8 \lambda_h  \lambda_{hk}  + 
8 \lambda_{k} \lambda_{hk} + 4 \lambda_{hk}^2  \ , 
\end{eqnarray}
\begin{eqnarray}
16 \pi^2 \beta_{g'} &=&  \frac 5 3  \left( \frac{41}{10} +1 \right)g'^3 
\nonumber \\
16 \pi^2 \beta_{g} &=&- \frac{19}{6} g^3 
 \nonumber \\
16 \pi^2 \beta_{g_3} &=& - 7 g_3^3  \ , 
\end{eqnarray}
\begin{equation} 
16 \pi^2 \beta_t = y_t \left\{ \frac 9 2 y_t^2 -  \left(\frac{17}{12} g'^2 + 
\frac 9 4 g^2 + 8 g_3^2 \right) \right \} \ .
\end{equation} 
Here $g_3,g,g'$ are the SM $SU(3)_C$, $SU(2)_L$  and $U(1)_Y$ 
gauge couplings, respectively, and we have neglected all the Yukawa couplings 
but  the top quark Yukawa, $y_t$. We have also neglected the $f_{ab}, g_{ab}$ couplings because 
for the range of singlet scalar masses that we 
consider ($\leq$ 2 TeV), they are are severely constrained by LFV and are much smaller than 1 except for some corners of the parameter space where some of them could be order one. For the analysis of the vacuum stability of the scalar potential $f_{ab}, g_{ab}$ are  subdominant, 
 specially in the region of large and negative mixed  scalar couplings required to accommodate 
the diphoton excess in Higgs decays. For smaller mixed scalar couplings, however, a more detailed analysis including all Yukawa couplings and taking also into account leading two-loop effects (as well as top quark mass uncertainties for the Higgs quartic coupling) should be carried out, which is beyond the scope of this work.

\section{Loop Functions for $H\rightarrow \gamma \gamma$ and $H\rightarrow Z \gamma$} \label{loop}

$\bullet\,$  Functions relevant for $H \rightarrow \gamma \gamma$: 
\begin{eqnarray}
A_0(x) &=& -x+x^2 \, f \left (\frac{1}{x}\right )
\\
A_{1/2}(x) &=& 2x+ 2x (1- x) \, f \left (\frac{1}{x}\right )
\\
A_{1}(x) &=& -2-3x-3x (2-x) \, f \left (\frac{1}{x}\right )
\label{eq:A}
\end{eqnarray}

$\bullet\, $  Functions relevant for $H \rightarrow Z \gamma$:
\begin{eqnarray}
A_0(x,y) &= &I_1(x,y)
\\
A_{1/2}(x,y)& = & I_1(x,y)-I_2(x,y) 
\\
A_{1}(x,y) &= & 4(3 - \tan^2 \theta_w) I_2(x,y) + 
\left[ (1 + 2 x^{-1}) \tan^2 \theta_w - (5 + 2 x^{-1})\right] I_1(x,y) 
\label{eq:AZG}
\end{eqnarray}
where 
\begin{eqnarray}
I_1(x,y) &=& \frac{x y}{2(x-y)} +  \frac{x^2 y^2}{2(x-y)^2} 
\left[f(x^{-1}) - f(y^{-1}) \right] + 
\frac{x^2 y}{(x-y)^2} \left[g(x^{-1}) - g(y^{-1}) \right] 
\\
I_2(x,y) &=& - \frac{x y}{2(x-y)} \left[f(x^{-1}) - f(y^{-1}) \right] 
\end{eqnarray}
and, for a Higgs mass below the kinematic threshold of the loop particle, $m_H < 2 m_{i}$, 
\begin{eqnarray}
f(x) &=& \arcsin^2 \sqrt{x} \,,
\\
g(x)&=& \sqrt{x^{-1} - 1} \arcsin \sqrt{x} \,.
\end{eqnarray}


\begin{thebibliography}{10}

\bibitem[1]{Mohapatra:2005wg}
R.~Mohapatra, S.~Antusch, K.~Babu, G.~Barenboim, M.-C. Chen, {\em et al.}, {\it
  {Theory of neutrinos: A White paper}},
  \href{http://dx.doi.org/10.1088/0034-4885/70/11/R02}{{\em Rept.Prog.Phys.}
  {\bf 70} (2007)  1757--1867},
  [\href{http://arxiv.org/abs/hep-ph/0510213}{{\tt arXiv:hep-ph/0510213}}]
[\href{http://inspirehep.net/record/695268}{{\tt InSPIRE}}].

\bibitem[2]{GonzalezGarcia:2007ib}
M.~Gonzalez-Garcia and M.~Maltoni, {\it {Phenomenology with Massive
  Neutrinos}},  \href{http://dx.doi.org/10.1016/j.physrep.2007.12.004}{{\em
  Phys.Rept.} {\bf 460} (2008)  1--129},
  [\href{http://arxiv.org/abs/0704.1800}{{\tt arXiv:0704.1800}}]
[\href{http://inspirehep.net/record/748589}{{\tt InSPIRE}}].

\bibitem[3]{Beringer:1900zz}
{\bf Particle Data Group}, J.~Beringer {\em et al.}, {\it {Review of Particle
  Physics (RPP)}},  \href{http://dx.doi.org/10.1103/PhysRevD.86.010001}{{\em
  Phys.Rev.} {\bf D86} (2012)  010001}
[\href{http://inspirehep.net/record/1126428}{{\tt InSPIRE}}].

\bibitem[4]{deGouvea:2013onf}
{\bf Intensity Frontier Neutrino Working Group}, A.~de~Gouvea {\em et al.},
  {\it {Neutrinos}},  [\href{http://arxiv.org/abs/1310.4340}{{\tt
  arXiv:1310.4340}}]
[\href{http://inspirehep.net/record/1260555}{{\tt InSPIRE}}].

\bibitem[5]{Zee:1985id}
A.~Zee, {\it {Quantum Numbers of Majorana Neutrino Masses}},
  \href{http://dx.doi.org/10.1016/0550-3213(86)90475-X}{{\em Nucl.Phys.} {\bf
  B264} (1986)  99}
[\href{http://inspirehep.net/record/218115}{{\tt InSPIRE}}].

\bibitem[6]{Babu:1988ki}
K.~Babu, {\it {Model of 'Calculable' Majorana Neutrino Masses}},
  \href{http://dx.doi.org/10.1016/0370-2693(88)91584-5}{{\em Phys.Lett.} {\bf
  B203} (1988)  132}
[\href{http://inspirehep.net/record/22952}{{\tt InSPIRE}}].

\bibitem[7]{Cheng:1980qt}
T.~Cheng and L.-F. Li, {\it {Neutrino Masses, Mixings and Oscillations in SU(2)
  x U(1) Models of Electroweak Interactions}},
  \href{http://dx.doi.org/10.1103/PhysRevD.22.2860}{{\em Phys.Rev.} {\bf D22}
  (1980)  2860}
[\href{http://inspirehep.net/record/9562}{{\tt InSPIRE}}].

\bibitem[8]{delAguila:2011gr}
F.~del Aguila, A.~Aparici, S.~Bhattacharya, A.~Santamaria, and J.~Wudka, {\it
  {A realistic model of neutrino masses with a large neutrinoless double beta
  decay rate}},  \href{http://dx.doi.org/10.1007/JHEP05(2012)133}{{\em JHEP}
  {\bf 1205} (2012)  133}, [\href{http://arxiv.org/abs/1111.6960}{{\tt
  arXiv:1111.6960}}]
[\href{http://inspirehep.net/record/1079241}{{\tt InSPIRE}}].

\bibitem[9]{delAguila:2013zba}
F.~del Aguila, A.~Aparici, S.~Bhattacharya, A.~Santamaria, and J.~Wudka, {\it
  {Neutrinoless double $\beta$ decay with small neutrino masses}},  {\em PoS}
  {\bf Corfu2012} (2013)  028, [\href{http://arxiv.org/abs/1305.4900}{{\tt
  arXiv:1305.4900}}]
[\href{http://inspirehep.net/record/1234427}{{\tt InSPIRE}}].

\bibitem[10]{Babu:2002uu}
K.~Babu and C.~Macesanu, {\it {Two loop neutrino mass generation and its
  experimental consequences}},
  \href{http://dx.doi.org/10.1103/PhysRevD.67.073010}{{\em Phys.Rev.} {\bf D67}
  (2003)  073010}, [\href{http://arxiv.org/abs/hep-ph/0212058}{{\tt
  arXiv:hep-ph/0212058}}]
[\href{http://inspirehep.net/record/603722}{{\tt InSPIRE}}].

\bibitem[11]{AristizabalSierra:2006gb}
D.~Aristizabal~Sierra and M.~Hirsch, {\it {Experimental tests for the Babu-Zee
  two-loop model of Majorana neutrino masses}},
  \href{http://dx.doi.org/10.1088/1126-6708/2006/12/052}{{\em JHEP} {\bf 0612}
  (2006)  052}, [\href{http://arxiv.org/abs/hep-ph/0609307}{{\tt
  arXiv:hep-ph/0609307}}]
[\href{http://inspirehep.net/record/727379}{{\tt InSPIRE}}].

\bibitem[12]{Nebot:2007bc}
M.~Nebot, J.~F. Oliver, D.~Palao, and A.~Santamaria, {\it {Prospects for the
  Zee-Babu Model at the CERN LHC and low energy experiments}},
  \href{http://dx.doi.org/10.1103/PhysRevD.77.093013}{{\em Phys.Rev.} {\bf D77}
  (2008)  093013}, [\href{http://arxiv.org/abs/0711.0483}{{\tt
  arXiv:0711.0483}}]
[\href{http://inspirehep.net/record/766680}{{\tt InSPIRE}}].

\bibitem[13]{Ohlsson:2009vk}
T.~Ohlsson, T.~Schwetz, and H.~Zhang, {\it {Non-standard neutrino interactions
  in the Zee-Babu model}},
  \href{http://dx.doi.org/10.1016/j.physletb.2009.10.025}{{\em Phys.Lett.} {\bf
  B681} (2009)  269--275}, [\href{http://arxiv.org/abs/0909.0455}{{\tt
  arXiv:0909.0455}}]
[\href{http://inspirehep.net/record/830191}{{\tt InSPIRE}}].

\bibitem[14]{Abe:2011fz}
{\bf DOUBLE-CHOOZ Collaboration}, Y.~Abe {\em et al.}, {\it {Indication for the
  disappearance of reactor electron antineutrinos in the Double Chooz
  experiment}},  \href{http://dx.doi.org/10.1103/PhysRevLett.108.131801}{{\em
  Phys.Rev.Lett.} {\bf 108} (2012)  131801},
  [\href{http://arxiv.org/abs/1112.6353}{{\tt arXiv:1112.6353}}]
[\href{http://inspirehep.net/record/1082938}{{\tt InSPIRE}}].

\bibitem[15]{An:2012eh}
{\bf DAYA-BAY Collaboration}, F.~An {\em et al.}, {\it {Observation of
  electron-antineutrino disappearance at Daya Bay}},
  \href{http://dx.doi.org/10.1103/PhysRevLett.108.171803}{{\em Phys.Rev.Lett.}
  {\bf 108} (2012)  171803}, [\href{http://arxiv.org/abs/1203.1669}{{\tt
  arXiv:1203.1669}}]
[\href{http://inspirehep.net/record/1093266}{{\tt InSPIRE}}].

\bibitem[16]{Ahn:2012nd}
{\bf RENO collaboration}, J.~Ahn {\em et al.}, {\it {Observation of Reactor
  Electron Antineutrino Disappearance in the RENO Experiment}},
  \href{http://dx.doi.org/10.1103/PhysRevLett.108.191802}{{\em Phys.Rev.Lett.}
  {\bf 108} (2012)  191802}, [\href{http://arxiv.org/abs/1204.0626}{{\tt
  arXiv:1204.0626}}]
[\href{http://inspirehep.net/record/1102875}{{\tt InSPIRE}}].

\bibitem[17]{Adam:2013mnn}
{\bf MEG Collaboration}, J.~Adam {\em et al.}, {\it {New constraint on the
  existence of the $\mu\rightarrow e \gamma$ decay}},
  \href{http://dx.doi.org/10.1103/PhysRevLett.110.201801}{{\em Phys.Rev.Lett.}
  {\bf 110} (2013)  201801}, [\href{http://arxiv.org/abs/1303.0754}{{\tt
  arXiv:1303.0754}}]
[\href{http://inspirehep.net/record/1222334}{{\tt InSPIRE}}].

\bibitem[18]{ATLAS:2012hi}
{\bf ATLAS Collaboration}, G.~Aad {\em et al.}, {\it {Search for doubly-charged
  Higgs bosons in like-sign dilepton final states at $\sqrt{s}=7$ TeV with the
  ATLAS detector}},
  \href{http://dx.doi.org/10.1140/epjc/s10052-012-2244-2}{{\em Eur.Phys.J.}
  {\bf C72} (2012)  2244}, [\href{http://arxiv.org/abs/1210.5070}{{\tt
  arXiv:1210.5070}}]
[\href{http://inspirehep.net/record/1191430}{{\tt InSPIRE}}].

\bibitem[19]{Chatrchyan:2012ya}
{\bf CMS Collaboration}, S.~Chatrchyan {\em et al.}, {\it {A search for a
  doubly-charged Higgs boson in $pp$ collisions at $\sqrt{s}=7$ TeV}},
  \href{http://dx.doi.org/10.1140/epjc/s10052-012-2189-5}{{\em Eur.Phys.J.}
  {\bf C72} (2012)  2189}, [\href{http://arxiv.org/abs/1207.2666}{{\tt
  arXiv:1207.2666}}]
[\href{http://inspirehep.net/record/1122035}{{\tt InSPIRE}}].

\bibitem[20]{Aad:2012tfa}
{\bf ATLAS Collaboration}, G.~Aad {\em et al.}, {\it {Observation of a new
  particle in the search for the Standard Model Higgs boson with the ATLAS
  detector at the LHC}},
  \href{http://dx.doi.org/10.1016/j.physletb.2012.08.020}{{\em Phys.Lett.} {\bf
  B716} (2012)  1--29}, [\href{http://arxiv.org/abs/1207.7214}{{\tt
  arXiv:1207.7214}}]
[\href{http://inspirehep.net/record/1124337}{{\tt InSPIRE}}].

\bibitem[21]{Chatrchyan:2012ufa}
{\bf CMS Collaboration}, S.~Chatrchyan {\em et al.}, {\it {Observation of a new
  boson at a mass of 125 GeV with the CMS experiment at the LHC}},
  \href{http://dx.doi.org/10.1016/j.physletb.2012.08.021}{{\em Phys.Lett.} {\bf
  B716} (2012)  30--61}, [\href{http://arxiv.org/abs/1207.7235}{{\tt
  arXiv:1207.7235}}]
[\href{http://inspirehep.net/record/1124338}{{\tt InSPIRE}}].

\bibitem[22]{Chao:2012xt}
W.~Chao, J.-H. Zhang, and Y.~Zhang, {\it {Vacuum Stability and Higgs Diphoton
  Decay Rate in the Zee-Babu Model}},
  \href{http://dx.doi.org/10.1007/JHEP06(2013)039}{{\em JHEP} {\bf 1306} (2013)
   039}, [\href{http://arxiv.org/abs/1212.6272}{{\tt arXiv:1212.6272}}]
[\href{http://inspirehep.net/record/1209025}{{\tt InSPIRE}}].

\bibitem[23]{Dicus:1992vj}
D.~Dicus and V.~Mathur, {\it {Upper bounds on the values of masses in unified
  gauge theories}},  \href{http://dx.doi.org/10.1103/PhysRevD.7.3111}{{\em
  Phys.Rev.} {\bf D7} (1973)  3111--3114}
[\href{http://inspirehep.net/record/334983}{{\tt InSPIRE}}].

\bibitem[24]{Lee:1977eg}
B.~W. Lee, C.~Quigg, and H.~Thacker, {\it {Weak Interactions at Very
  High-Energies: The Role of the Higgs Boson Mass}},
  \href{http://dx.doi.org/10.1103/PhysRevD.16.1519}{{\em Phys.Rev.} {\bf D16}
  (1977)  1519}
[\href{http://inspirehep.net/record/119348}{{\tt InSPIRE}}].

\bibitem[25]{Frere:1983ag}
J.~Frere, D.~Jones, and S.~Raby, {\it {Fermion Masses and Induction of the Weak
  Scale by Supergravity}},
  \href{http://dx.doi.org/10.1016/0550-3213(83)90606-5}{{\em Nucl.Phys.} {\bf
  B222} (1983)  11}
[\href{http://inspirehep.net/record/12950}{{\tt InSPIRE}}].

\bibitem[26]{AlvarezGaume:1983gj}
L.~Alvarez-Gaume, J.~Polchinski, and M.~B. Wise, {\it {Minimal Low-Energy
  Supergravity}},  \href{http://dx.doi.org/10.1016/0550-3213(83)90591-6}{{\em
  Nucl.Phys.} {\bf B221} (1983)  495}
[\href{http://inspirehep.net/record/189012}{{\tt InSPIRE}}].

\bibitem[27]{Casas:1996de}
J.~Casas and S.~Dimopoulos, {\it {Stability bounds on flavor violating
  trilinear soft terms in the MSSM}},
  \href{http://dx.doi.org/10.1016/0370-2693(96)01000-3}{{\em Phys.Lett.} {\bf
  B387} (1996)  107--112}, [\href{http://arxiv.org/abs/hep-ph/9606237}{{\tt
  arXiv:hep-ph/9606237}}]
[\href{http://inspirehep.net/record/419309}{{\tt InSPIRE}}].

\bibitem[28]{McDonald:2003zj}
K.~L. McDonald and B.~McKellar, {\it {Evaluating the two loop diagram
  responsible for neutrino mass in Babu's model}},
  [\href{http://arxiv.org/abs/hep-ph/0309270}{{\tt arXiv:hep-ph/0309270}}]
[\href{http://inspirehep.net/record/628953}{{\tt InSPIRE}}].

\bibitem[29]{Pich:2013lsa}
A.~Pich, {\it {Precision Tau Physics}},
  [\href{http://arxiv.org/abs/1310.7922}{{\tt arXiv:1310.7922}}]
[\href{http://inspirehep.net/record/1262571}{{\tt InSPIRE}}].

\bibitem[30]{Raidal:1997hq}
M.~Raidal and A.~Santamaria, {\it {Muon electron conversion in nuclei versus
  $mu \rightarrow e \gamma$: an effective field theory point of view}},
  \href{http://dx.doi.org/10.1016/S0370-2693(98)00020-3}{{\em Phys.Lett.} {\bf
  B421} (1998)  250--258}, [\href{http://arxiv.org/abs/hep-ph/9710389}{{\tt
  arXiv:hep-ph/9710389}}]
[\href{http://inspirehep.net/record/449901}{{\tt InSPIRE}}].

\bibitem[31]{Witte:2012zza}
H.~Witte, B.~Muratori, K.~Hock, R.~Appleby, H.~Owen, {\em et al.}, {\it {Status
  of the PRISM FFAG Design for the Next Generation Muon-to-Electron Conversion
  Experiment}},  {\em Conf.Proc.} {\bf C1205201} (2012)  79--81
[\href{http://inspirehep.net/record/1125474}{{\tt InSPIRE}}].

\bibitem[32]{Kuno:2013mha}
{\bf COMET Collaboration}, Y.~Kuno, {\it {A search for muon-to-electron
  conversion at J-PARC: The COMET experiment}},
  \href{http://dx.doi.org/10.1093/ptep/pts089}{{\em PTEP} {\bf 2013} (2013)
  022C01}
[\href{http://inspirehep.net/record/1223754}{{\tt InSPIRE}}].

\bibitem[33]{Onorato:2013uka}
{\bf mu2e Collaboration}, G.~Onorato, {\it {The Mu2e experiment at Fermilab:
  $\mu^{-} N \to e^{-} N$}},
  \href{http://dx.doi.org/10.1016/j.nima.2012.11.073}{{\em Nucl.Instrum.Meth.}
  {\bf A718} (2013)  102--103}
[\href{http://inspirehep.net/record/1250204}{{\tt InSPIRE}}].

\bibitem[34]{delAguila:2012nu}
F.~del Aguila, A.~Aparici, S.~Bhattacharya, A.~Santamaria, and J.~Wudka, {\it
  {Effective Lagrangian approach to neutrinoless double beta decay and neutrino
  masses}},  \href{http://dx.doi.org/10.1007/JHEP06(2012)146}{{\em JHEP} {\bf
  1206} (2012)  146}, [\href{http://arxiv.org/abs/1204.5986}{{\tt
  arXiv:1204.5986}}]
[\href{http://inspirehep.net/record/1112579}{{\tt InSPIRE}}].

\bibitem[35]{Tang:2009na}
J.~Tang and W.~Winter, {\it {Physics with near detectors at a neutrino
  factory}},  \href{http://dx.doi.org/10.1103/PhysRevD.80.053001}{{\em
  Phys.Rev.} {\bf D80} (2009)  053001},
  [\href{http://arxiv.org/abs/0903.3039}{{\tt arXiv:0903.3039}}]
[\href{http://inspirehep.net/record/815682}{{\tt InSPIRE}}].

\bibitem[36]{delAguila:2013mia}
F.~del Aguila and M.~Chala, {\it {LHC bounds on Lepton Number Violation
  mediated by doubly and singly-charged scalars}},
  [\href{http://arxiv.org/abs/1311.1510}{{\tt arXiv:1311.1510}}]
[\href{http://inspirehep.net/record/1263664}{{\tt InSPIRE}}].

\bibitem[37]{Sugiyama:2012yw}
H.~Sugiyama, K.~Tsumura, and H.~Yokoya, {\it {Discrimination of models
  including doubly charged scalar bosons by using tau lepton decay
  distributions}},
  \href{http://dx.doi.org/10.1016/j.physletb.2012.09.044}{{\em Phys.Lett.} {\bf
  B717} (2012)  229--234}, [\href{http://arxiv.org/abs/1207.0179}{{\tt
  arXiv:1207.0179}}]
[\href{http://inspirehep.net/record/1120759}{{\tt InSPIRE}}].

\bibitem[38]{delAguila:2013yaa}
F.~del Aguila, M.~Chala, A.~Santamaria, and J.~Wudka, {\it {Discriminating
  between lepton number violating scalars using events with four and three
  charged leptons at the LHC}},
  \href{http://dx.doi.org/10.1016/j.physletb.2013.07.014}{{\em Phys.Lett.} {\bf
  B725} (2013)  310--315}, [\href{http://arxiv.org/abs/1305.3904}{{\tt
  arXiv:1305.3904}}]
[\href{http://inspirehep.net/record/1233724}{{\tt InSPIRE}}].

\bibitem[39]{delAguila:2013hla}
F.~del Aguila, M.~Chala, A.~Santamaria, and J.~Wudka, {\it {Distinguishing
  between lepton number violating scalars at the LHC}},
  [\href{http://arxiv.org/abs/1307.0510}{{\tt arXiv:1307.0510}}]
[\href{http://inspirehep.net/record/1241429}{{\tt InSPIRE}}].

\bibitem[40]{Muhlleitner:2003me}
M.~Muhlleitner and M.~Spira, {\it {A Note on doubly charged Higgs pair
  production at hadron colliders}},
  \href{http://dx.doi.org/10.1103/PhysRevD.68.117701}{{\em Phys.Rev.} {\bf D68}
  (2003)  117701}, [\href{http://arxiv.org/abs/hep-ph/0305288}{{\tt
  arXiv:hep-ph/0305288}}]
[\href{http://inspirehep.net/record/619556}{{\tt InSPIRE}}].

\bibitem[41]{Kanemura:2000bq}
S.~Kanemura, T.~Kasai, G.-L. Lin, Y.~Okada, J.-J. Tseng, {\em et al.}, {\it
  {Phenomenology of Higgs bosons in the Zee model}},
  \href{http://dx.doi.org/10.1103/PhysRevD.64.053007}{{\em Phys.Rev.} {\bf D64}
  (2001)  053007}, [\href{http://arxiv.org/abs/hep-ph/0011357}{{\tt
  arXiv:hep-ph/0011357}}]
[\href{http://inspirehep.net/record/537580}{{\tt InSPIRE}}].

\bibitem[42]{Kannike:2012pe}
K.~Kannike, {\it {Vacuum Stability Conditions From Copositivity Criteria}},
  \href{http://dx.doi.org/10.1140/epjc/s10052-012-2093-z}{{\em Eur.Phys.J.}
  {\bf C72} (2012)  2093}, [\href{http://arxiv.org/abs/1205.3781}{{\tt
  arXiv:1205.3781}}]
[\href{http://inspirehep.net/record/1115191}{{\tt InSPIRE}}].

\bibitem[43]{Degrassi:2012ry}
G.~Degrassi, S.~Di~Vita, J.~Elias-Miro, J.~R. Espinosa, G.~F. Giudice, {\em et
  al.}, {\it {Higgs mass and vacuum stability in the Standard Model at NNLO}},
  \href{http://dx.doi.org/10.1007/JHEP08(2012)098}{{\em JHEP} {\bf 1208} (2012)
   098}, [\href{http://arxiv.org/abs/1205.6497}{{\tt arXiv:1205.6497}}]
[\href{http://inspirehep.net/record/1116539}{{\tt InSPIRE}}].

\bibitem[44]{ATLAS:2013oma}
{\bf ATLAS Collaboration}, {\it {Measurements of the properties of the
  Higgs-like boson in the two photon decay channel with the ATLAS detector
  using 25 $\mathrm{fb}^{-1}$ of proton-proton collision data}},
[\href{http://inspirehep.net/record/1229970}{{\tt InSPIRE}}].

\bibitem[45]{CMS:ril}
{\bf CMS Collaboration}, {\it {Updated measurements of the Higgs boson at 125
  GeV in the two photon decay channel}},
[\href{http://inspirehep.net/record/1230222}{{\tt InSPIRE}}].

\bibitem[46]{Carena:2012xa}
M.~Carena, I.~Low, and C.~E. Wagner, {\it {Implications of a Modified Higgs to
  Diphoton Decay Width}},
  \href{http://dx.doi.org/10.1007/JHEP08(2012)060}{{\em JHEP} {\bf 1208} (2012)
   060}, [\href{http://arxiv.org/abs/1206.1082}{{\tt arXiv:1206.1082}}]
[\href{http://inspirehep.net/record/1117441}{{\tt InSPIRE}}].

\bibitem[47]{Ellis:1975ap}
J.~R. Ellis, M.~K. Gaillard, and D.~V. Nanopoulos, {\it {A Phenomenological
  Profile of the Higgs Boson}},
  \href{http://dx.doi.org/10.1016/0550-3213(76)90382-5}{{\em Nucl.Phys.} {\bf
  B106} (1976)  292}
[\href{http://inspirehep.net/record/100355}{{\tt InSPIRE}}].

\bibitem[48]{Shifman:1979eb}
M.~A. Shifman, A.~Vainshtein, M.~Voloshin, and V.~I. Zakharov, {\it {Low-Energy
  Theorems for Higgs Boson Couplings to Photons}},  {\em Sov.J.Nucl.Phys.} {\bf
  30} (1979)  711--716
[\href{http://inspirehep.net/record/141287}{{\tt InSPIRE}}].

\bibitem[49]{Gunion:1989we}
J.~F. Gunion, H.~E. Haber, G.~L. Kane, and S.~Dawson, {\it {The Higgs Hunter's
  Guide}},  {\em Front.Phys.} {\bf 80} (2000)  1--448
[\href{http://inspirehep.net/record/279039}{{\tt InSPIRE}}].

\bibitem[50]{Chen:2013vi}
C.-S. Chen, C.-Q. Geng, D.~Huang, and L.-H. Tsai, {\it {New Scalar
  Contributions to $h\to Z\gamma$}},
  \href{http://dx.doi.org/10.1103/PhysRevD.87.075019}{{\em Phys.Rev.} {\bf D87}
  (2013)  075019}, [\href{http://arxiv.org/abs/1301.4694}{{\tt
  arXiv:1301.4694}}]
[\href{http://inspirehep.net/record/1215316}{{\tt InSPIRE}}].

\bibitem[51]{GonzalezGarcia:2012sz}
M.~Gonzalez-Garcia, M.~Maltoni, J.~Salvado, and T.~Schwetz, {\it {Global fit to
  three neutrino mixing: critical look at present precision}},
  \href{http://dx.doi.org/10.1007/JHEP12(2012)123}{{\em JHEP} {\bf 1212} (2012)
   123}, [\href{http://arxiv.org/abs/1209.3023}{{\tt arXiv:1209.3023}}]
[\href{http://inspirehep.net/record/1185570}{{\tt InSPIRE}}].

\bibitem[52]{Tortola:2012te}
D.~Forero, M.~Tortola, and J.~Valle, {\it {Global status of neutrino
  oscillation parameters after Neutrino-2012}},
  \href{http://dx.doi.org/10.1103/PhysRevD.86.073012}{{\em Phys.Rev.} {\bf D86}
  (2012)  073012}, [\href{http://arxiv.org/abs/1205.4018}{{\tt
  arXiv:1205.4018}}]
[\href{http://inspirehep.net/record/1115188}{{\tt InSPIRE}}].

\bibitem[53]{Fogli:2012ua}
G.~Fogli, E.~Lisi, A.~Marrone, D.~Montanino, A.~Palazzo, {\em et al.}, {\it
  {Global analysis of neutrino masses, mixings and phases: entering the era of
  leptonic CP violation searches}},
  \href{http://dx.doi.org/10.1103/PhysRevD.86.013012}{{\em Phys.Rev.} {\bf D86}
  (2012)  013012}, [\href{http://arxiv.org/abs/1205.5254}{{\tt
  arXiv:1205.5254}}]
[\href{http://inspirehep.net/record/1115710}{{\tt InSPIRE}}].

\bibitem[54]{Schmidt:2014zoa}
D.~Schmidt, T.~Schwetz, and H.~Zhang, {\it {Status of the Zee-Babu model for
  neutrino mass and possible tests at a like-sign linear collider}},
  [\href{http://arxiv.org/abs/1402.2251}{{\tt arXiv:1402.2251}}]
[\href{http://inspirehep.net/record/1280733}{{\tt InSPIRE}}].

\end{thebibliography}

\providecommand{\href}[2]{#2}\begingroup\raggedright\endgroup

\end{document}